\documentclass[aps,prd,reprint,superscriptaddress,nofootinbib]{revtex4-1}

\usepackage{hyperref}
\hypersetup{colorlinks=true}
\usepackage{lipsum}
\usepackage{amsmath}
\usepackage{amssymb}
\usepackage{tensor}
\usepackage{graphicx}
\usepackage[utf8]{inputenc}
\usepackage{multirow} 
\usepackage{mathrsfs} 

\newcommand{\order}{\mathcal{O}}
\newcommand{\bs}{\boldsymbol}
\newcommand{\mb}{\mathbf}

\newcommand{\doe}{\partial}
\newcommand{\mc}{\mathcal}

\begin{document}

\title{Gravitational waves from spinning binary black holes\\ at the leading post-Newtonian orders at all orders in spin}
\author{Nils Siemonsen}
\affiliation{Institut für Physik, Humboldt-Universität zu Berlin, Newtonstraße 15, D-12489 Berlin, Germany}
\author{Jan Steinhoff}
\affiliation{Max Planck Institute for Gravitational Physics (Albert Einstein Institute), Am Mühlenberg 1, Potsdam D-14476, Germany}
\author{Justin Vines}
\affiliation{Max Planck Institute for Gravitational Physics (Albert Einstein Institute), Am Mühlenberg 1, Potsdam D-14476, Germany}

\date{\today}

\begin{abstract}
We determine the binding energy, the total gravitational wave energy flux, and the gravitational wave modes for a binary of rapidly spinning black holes, working in linearized gravity and at leading orders in the orbital velocity, but to all orders in the black holes' spins. 
Though the spins are treated nonperturbatively, surprisingly, the binding energy and the flux are given by simple analytical expressions which are finite (respectively third- and fifth-order) polynomials in the spins. Our final results are restricted to the important case of quasicircular orbits with the black holes' spins aligned with the orbital angular momentum.
\end{abstract}


\maketitle

\section{Introduction}

The general relativistic two-body problem, in particular the description of compact binaries as one of the most important sources of gravitational waves (GWs) detectable from Earth, poses a large challenge for analytical as well as numerical calculations.  Accurate waveform models and templates generated by such calculations are utilized in matched filtering techniques to confidently assign gravitational wave signals to compact binaries, extract information about the source, and investigate possible deviations from the predictions of general relativity (GR). Recent detections of such signals \cite{Abbott:2016nmj,Abbott:2016blz,Abbott:2017oio,TheLIGOScientific:2017qsa,Abbott:2017vtc,Abbott:2017gyy,Abbott:2017xzu} have demonstrated the success of these approaches, but the need for more accurate and more general descriptions of binary dynamics in GR persists and will grow with future more sensitive GW detectors.

Several analytical approaches to the two-body problem in GR have been developed over the past century. The predominant method for describing the dynamics and predicting the form of the emitted radiation for arbitrary-mass-ratio compact binaries employs the post-Newtonian (PN) approximation \cite{Blanchet:2013haa,Futamase:2007zz}. The PN approach seeks to extract information from full GR by perturbatively expanding the dynamics and radiation about the slow-motion, weak-field regime.  Applied to a binary system, the computed dynamics and waveforms are expected to provide accurate predictions only in the early inspiral phase. However, analytic results from the PN approximation, together with numerical descriptions of the late inspiral, plunge, and merger, can be combined synergetically in effective-one-body (EOB) models \cite{Buonanno:1998gg,Buonanno:2000ef}. In such, the binary's evolution can be accurately modeled from the early inspiral stage through the merger and ringdown of the remnant. In order for EOB models to efficiently extract information from detected gravitational wave signals, accurate analytical and numerical input must be provided. Thus, efforts continue to push PN theory to ever increasing accuracy. 

The PN approximation is an expansion of a compact binary's full general relativistic dynamics in the dimensionless small parameter $\epsilon_{\text{PN}}\sim v^2 / c^2 \sim Gm /c^2 r$.\footnote{As usual, $G$ is Newton's constant, $c$ the speed of light, $m$ the binary's mass scale, $v$ its orbital velocity, and $r$ the orbital separation.} This amounts to using $1/c^2$ as the formal expansion parameter; an $\order(c^{-2n})$ contribution is said to be at $n$PN order.  The approximation has been pushed to 4PN order in the conservative sector \cite{Damour:2014jta, Bernard:2016wrg} and to 3PN for the gravitational wave modes \cite{Blanchet:2008je, Blanchet:2013haa} for binaries with constituents represented as structureless (monopole) point particles, which could describe nonspinning binary black holes (BBHs). In the seminal works of Mathisson, Papapetrou \cite{Mathisson:1937zz,Mathisson:2010,Papapetrou:1951pa}, and later Dixon \cite{Dixon:1979, Dixon:2015vxa}, finite-size effects for an extended body in GR have been modeled in the form of a multipolar structure associated with the body. Subsequently, spin is incorporated in PN theory through spin-dependent multipole moments attached to a point-particle representation of the body.  Spin, besides the field strength and velocity, is typically treated perturbatively in PN calculations, expanding in the dimensionless small parameter $\epsilon_{\text{spin}}\sim Gm\chi/rc^2$.\footnote{Here, $\chi=cS/Gm^2$ is the dimensionless spin parameter for a body of mass $m$ and spin $S$, with $\chi\in [0,1)$ for a black hole.} For rapidly rotating black holes, one finds that $\epsilon_{\text{spin}}\sim\epsilon_{\text{PN}}$. Previously, the conservative dynamics of spinning BBHs was computed up to 4PN order \cite{Levi:2016ofk}; for the leading orders at various orders in spin see Refs.~\cite{Tulczyjew:1959b, Barker:1970zr, Barker:1975ae, D'Eath:1975vw, Thorne:1984mz, Poisson:1997ha, Damour:2001tu, Hergt:2007ha, Hergt:2008jn, Levi:2014gsa, Vaidya:2014kza, Marsat:2014xea, Vines:2016qwa}. For spin-dependent leading orders in the radiative sector, see Refs.~\cite{Kidder:1992fr, Kidder:1995zr, Poisson:1997ha, Buonanno:2012rv, Marsat:2014xea, Mikoczi:2005dn}. The traditional approach has been to truncate the expansion in $\epsilon_{\text{spin}}$ in accord with the order counting made natural by the relation $\epsilon_{\text{spin}}\sim\epsilon_{\text{PN}}$.  At 4PN order, this corresponds to a truncation at fourth order in the spins.

In this paper, we treat $\epsilon_{\text{PN}}$ and $\epsilon_{\text{spin}}$ as independent expansion parameters, and we determine the conservative and radiative dynamics of an arbitrary-mass-ratio BBH at leading PN order (leading order in $\epsilon_{\text{PN}}$), but to \textit{all orders in spin} (no expansion in $\epsilon_{\text{spin}}$) \cite{Vines:2016qwa}. Therefore, we obtain terms which are, according to the traditional PN counting in $1/c^2$, of \textit{arbitrarily large PN order}. Our final results are valid for circular orbits and spins aligned with the orbital angular momentum (though several intermediate results are applicable to more general situations). This is an important case since waveforms for aligned (nonprecessing) spins can approximate \emph{precessing} waveforms very well when viewed from a certain frame \cite{Pan:2013rra} and hence can form the basis for precessing waveform models. Our results include the leading-PN-order interactions with spin powers $S_1^{n_1} S_2^{n_2}$ for all combinations of non-negative integers $n_1$, $n_2$ (covering both ``$S_1$-$S_2$'' and ``self-spin'' interactions) in the form of compact expressions, which are resummed from the perspective of the traditional PN counting. This nonperturbative aspect of our results can improve the synergetic EOB waveform model by suggesting improved resummations of perturbative PN results in future work.

In the framework of an effective action principle, we model the BBH as two point particles with infinite sets of spin-induced multipole moments, fixed by matching to the Kerr metric. We consider the coupling of this multipolar structure to gravity at first post-Minkowskian (PM) order (i.e.\ in linearized gravity), in Sec.~\ref{sectioneffectiveactionprinciple}. In Sec.~\ref{sectionconservativedynamics}, the near zone field equations are solved at leading post-Newtonian order (at leading orders in the orbital velocity), and the binding energy of the binary is computed for circular orbits and spins aligned with the orbital angular momentum. Lastly, for the same orbital configuration and at the same level of approximation, the source multipole moments of the complete system are computed and employed to determine the GW modes and total GW energy flux emitted by the BBH in Sec.~\ref{sectionfluxandmodes}.

Throughout the paper, Greek letters $\mu, \nu, \alpha, \beta,...$ are used as spacetime (abstract or coordinate-basis) indices.  After Sec.~\ref{sectioneffectiveactionprinciple}, Latin letters $i, j, k, a, b,...$ are used as spatial indices.  Various other types of Latin indices are used as indicated in the text.  We exploit the multi-index notation $L:=\mu_1\dots \mu_\ell$ for some multi-index tensors, or for $\ell$ tensorial powers of a vector $v^\mu$ such that $v^L=v^{\mu_1\dots \mu_\ell}:=v^{\mu_1}\dots v^{\mu_\ell}$, as well as $L-1=\mu_l \dots \mu_{\ell-1}$, etc., both for spacetime indices as here and for spatial indices, $L=i_1\ldots i_\ell$, the distinction being clear from the context.  Our sign convention for the volume form is such that $\epsilon_{0123}=+1$ in a local Minkowskian basis, and our sign convention for the Riemann tensor is such that $2\nabla_{[\mu}\nabla_{\nu]}w_\alpha=R_{\mu\nu\alpha}{}^\beta w_\beta$.

\section{Effective action for a spinning black hole} \label{sectioneffectiveactionprinciple}

In this section, we review the construction of an effective action functional for a localized spinning body coupled to gravity, assuming the body has only translational and rotational degrees of freedom and only spin-induced multipole moments, as is appropriate for a spinning black hole at the orders considered in this paper.  We begin in Sec.~\ref{sec:effS} with a general such spinning body, seeing how its translational and rotational kinematics are linked to its universal (i.e., body-independent) monopole and dipole couplings to gravity.  In Sec.~\ref{sec:Lc}, we consider the couplings of the body's higher-order spin-induced multipoles (quadrupole, octupole, etc.) to the spacetime curvature, working at linear order in the curvature, and then specialize to the multipole structure of a spinning (Kerr) black hole.  While those two subsections work in what could be (in principle) a general curved spacetime, maintaining general covariance, we specialize to the case of first-post-Minkowskian spacetime (a linear perturbation of flat spacetime) in Sec.~\ref{sec:linapp}.

\subsection{Effective action of a spinning point-particle}\label{sec:effS}
 
The effective translational and rotational degrees of freedom for a spinning point particle can be taken to be a worldline $x=z (\lambda)$ depending on a generic parameter $\lambda$, with tangent $u^\mu=dz^\mu/d\lambda$, and 
a ``body-fixed'' tetrad $\tensor{\varepsilon}{_A^\mu}(\lambda)$ (defined only along the worldline) with orthonormality conditions:
\begin{align}
\tensor{\varepsilon}{_A^\mu}\tensor{\varepsilon}{^A^\nu}=g^{\mu\nu},  &\qquad \tensor{\varepsilon}{_A_\mu}\tensor{\varepsilon}{_B^\mu}=\eta_{AB},
\end{align} 
where $\eta_{AB}$ is a frame Minkowski metric, and $g^{\mu\nu}$ is the spacetime (inverse) metric (evaluated at $x=z$) which is used to raise or lower all spacetime indices in this subsection and the next.  For later convenience, we also define throughout spacetime a ``global'' tetrad field $\tensor{e}{_a^\mu}$ with analogous orthonormality conditions.  In this section, Latin letters $a, b$ are indices for the the local Lorentz basis at each spacetime point, and $A,B$ are indices for the body-fixed Lorentz basis along the worldline.  The two tetrads are related by a local Lorentz transformation at each point on the worldline: $\tensor{\varepsilon}{_A^\mu}=\tensor{\Lambda}{_A^a}\tensor{e}{_a^\mu}$. The angular velocity tensor,  
\begin{align}\label{Omega}
\Omega^{\mu\nu}=\tensor{\varepsilon}{_A^\mu}\frac{D\tensor{\varepsilon}{^A^\nu}}{d\lambda}, 
\end{align} 
with $\Omega^{\mu\nu}=-\Omega^{\nu\mu}$, serves as a measure of the particle's rotation along the worldline.
Of the six (local Lorentz transformation) degrees of freedom in the body-fixed tetrad $\tensor{\varepsilon}{_A^\mu}$, we expect three to be physical, describing the body's (spatial) rotation, while the other three (boosts) are redundant with the translational degrees of freedom.  We can remove this redundancy by fixing the timelike vector $\tensor{\varepsilon}{_0^\mu}$ to be the direction of the worldline tangent or velocity, 
\begin{align}\label{tetradconstraint}
\tensor{\varepsilon}{_0^\mu}=U^\mu:=u^\mu /\sqrt{-u_\rho u^\rho}.
\end{align}
This constraint on the tetrad will translate into a corresponding constraint on the body's spin tensor.

An effective action for the spinning point particle can be given as a functional of the worldline $z(\lambda)$, the body-fixed tetrad $\tensor{\varepsilon}{_A^\mu}(\lambda)$, and the spin tensor $S_{\mu\nu}(\lambda)$ conjugate to the angular velocity $\Omega^{\mu\nu}(\lambda)$ (in the sense of a Legendre transform) by \cite{Levi:2015msa,Porto:2016pyg,Hanson:1974qy}
\begin{align}
S_{\text{p.p.}}[z,\varepsilon,S]=\int d\lambda\, \bigg\{ - m \sqrt{-u_\mu u^\mu} +\frac{1}{2}S_{\mu\nu}\Omega^{\mu\nu} + \mathcal{L}_c \bigg\},
\label{poledipoleactiongeneral} 
\end{align} 
where $m$ is the conserved (bare) rest mass.  Here, the first two terms serve as translational and rotational ``kinetic terms,'' and they also implicitly encode the monopole- and dipole-type couplings to the spacetime geometry through $u_\mu u^\mu=g_{\mu\nu}u^\mu u^\nu$ and through the covariant derivative in (\ref{Omega}), respectively; the dipole couplings are explicitly extracted at the end of this subsection.  The couplings between the body's higher multipoles and the spacetime curvature are contained in $\mathcal{L}_c(z,U,S)$ as specified in the following subsection. 

The constraint (\ref{tetradconstraint}) on the tetrad translates \cite{Levi:2015msa,steinhoff15} into the following constraint on the spin tensor,
\begin{align}\label{covSSC}
S^{\mu\nu}U_\nu=0,
\end{align}
serving as a spin supplementary condition (SSC) \cite{Frenkel:1926zz,Mathisson:1937zz,Mathisson:2010}, which we will refer to as the \emph{covariant SSC}.\footnote{At the order considered in this paper, this SSC is equivalent to the condition $S^{\mu\nu} p_\nu = 0$ due to Tulczyjew, where $p_\nu$ is the linear momentum appearing in the Mathisson-Papapetrou-Dixon (MPD) equations \cite{Mathisson:1937zz,Mathisson:2010,Papapetrou:1951pa,Dixon:1979, Dixon:2015vxa}, which in general leads to a better behaved motion of the center \cite{Tulczyjew:1959,Kyrian:2007zz}.  The time evolution defined by the action does not automatically preserve the SSC. One can either incorporate the SSC into the action with a Lagrange multiplier, or insert a solution of the SSC into the action.}
In order to simplify our computations below, we implement a change of variables from the rotational degrees of freedom as appearing in the action (\ref{poledipoleactiongeneral}).
As in \cite{Levi:2015msa,steinhoff15,Vines:2016unv}, we apply a local Lorentz transformation to the body-fixed tetrad,
\begin{align}\label{boosttetrad}
\tensor{\varepsilon}{_A^\nu}\to\tensor{\tilde{\varepsilon}}{_A^\mu}=\tensor{L}{^\mu_\nu}\tensor{\varepsilon}{_A^\nu},
\end{align}
such that 
the timelike vector $\tensor{\varepsilon}{_0^\mu} = U^\mu$ as in (\ref{tetradconstraint}) is boosted into the direction of the timelike vector $\tensor{e}{_0^\mu}$ of the global tetrad: $\tensor{\tilde{\varepsilon}}{_0^\mu}=\tensor{L}{^\mu_\nu}\tensor{\varepsilon}{_0^\nu}=\tensor{e}{_0^\mu}$. This is accomplished with the standard boost
\begin{align}\label{boost}
\tensor{L}{^\mu_\nu}:=\delta^\mu_\nu-2\tensor{e}{_0^\mu}U_\nu+\frac{\omega^\mu \omega_\nu}{-U_\rho \omega^\rho}, & & \omega^\mu:=U^\mu +\tensor{e}{_0^\mu}.
\end{align} 
Under (\ref{boosttetrad}) with (\ref{boost}), the second term of (\ref{poledipoleactiongeneral}) becomes \cite{Levi:2015msa} 
\begin{align} 
\frac{1}{2}S_{\mu\nu}\Omega^{\mu\nu}=\frac{1}{2}\tilde{S}_{\mu\nu}\tilde{\Omega}^{\mu\nu}+\tilde{S}_{\mu\nu}U^\mu \frac{D U^\nu}{d\lambda},
\label{spintensorshift} 
\end{align}
where $\tilde\Omega^{\mu\nu}=\tilde\varepsilon_A{}^\mu (D\tilde\varepsilon^{A\nu}/d\lambda)$, and where the new spin tensor $\tilde{S}^{\mu\nu}$ is defined by 
\begin{align}\label{PPS}
S^{\mu\nu} &= \mathcal{P}^\mu_\alpha \mathcal{P}^\nu_\beta \tilde{S}^{\alpha\beta},
\\\label{NWSSC}
0&=\tilde{S}^{\mu\nu} (U_\nu + e_{0\nu}),
\end{align}
with
\begin{align}
\mathcal{P}^\mu_\nu := \delta^\mu_\nu + U^\mu U_\nu
\end{align}
 being the projector orthogonal to $U^\mu$.\footnote{\label{footNWSSC}As a result of the boost \eqref{boosttetrad}--\eqref{boost}, the covariant SSC \eqref{covSSC} for the original spin tensor $S^{\mu\nu}$ translates into the SSC for the new spin tensor $\tilde S^{\mu\nu}$ given by \eqref{NWSSC}, which (at the considered order) is (equivalent to) a \emph{canonical} or Newton-Wigner SSC \cite{Pryce:1935, Pryce:1948, Newton:Wigner:1949}.  In the setting of the MPD equations, the choice of the SSC, constraining the spin tensor, is linked to the choice of the body's central worldline $z$; the canonical SSC would define a new worldline $\tilde z$, differing from the covariant-SSC worldline $z$, and together $\tilde z$ and $\tilde S^{\mu\nu}$ would satisfy the MPD equations with \eqref{NWSSC}, as $z$ and $S^{\mu\nu}$ satisfy the MPD equations with \eqref{covSSC}.  In our treatment here, in order to simplify the effective action, we will use mixed variables, namely the covariant-SSC worldline $z$ and the canonical-SSC spin tensor $\tilde S^{\mu\nu}$, which do not satisfy the MPD equations.}

We can explicitly extract the dipole coupling to the spacetime geometry by switching from the coordinate basis to the global tetrad basis and from covariant derivatives to ordinary derivatives, recalling $\tensor{\varepsilon}{_A^\mu}=\tensor{\Lambda}{_A^a}\tensor{e}{_a^\mu}$,
\begin{align} 
\tilde{S}_{\mu\nu}\tilde{\Omega}^{\mu\nu}&=\tilde{S}_{ab}\tensor{\tilde{\Lambda}}{_A^a}\frac{d\tensor{\tilde{\Lambda}}{^A^b}}{d\lambda}+\omega_\mu{}^{ab} \tilde{S}_{ab}u^\mu, 
\label{spinconnectionterm} \\
\frac{D U^\nu}{d\lambda} &= \left[ \frac{d U^a}{d\lambda} + \omega_\beta{}^{ba} U_b u^\beta \right] \tensor{e}{_a^\nu} ,\label{DU}
\end{align}
where $\omega_\mu{}^{ab} = \tensor{e}{^b_\alpha}\nabla_\mu\tensor{e}{^a^\alpha} = \tensor{e}{^b_\alpha} \partial_\mu\tensor{e}{^a^\alpha} + \tensor{e}{^b_\alpha} \tensor{\Gamma}{^\alpha_\mu_\beta} \tensor{e}{^a^\beta}$ are the Ricci rotation coefficients for the global tetrad.  Finally, using \eqref{spintensorshift}--\eqref{DU}, the action \eqref{poledipoleactiongeneral} reads
\begin{align}
\begin{aligned} 
S_{\text{p.p.}}= & \int d\lambda\,\bigg\{ -m\sqrt{-u_\mu u^\mu}+\frac{1}{2}\tilde{S}_{ab}\tensor{\tilde{\Lambda}}{_A^a}\frac{d\tensor{\tilde{\Lambda}}{^A^b}}{d\lambda}\\ 
&+\tilde{S}_{ab}U^a \frac{dU^b}{d\lambda}+\frac{1}{2}\omega_\mu{}^{ab} S_{ab}u^\mu +\mathcal{L}_c\bigg\}. 
\end{aligned} 
\label{poledipoleactiongeneral2} 
\end{align} 
Starting from the following subsection, we drop the tildes used here to indicate the boosted variables; the use of the unboosted variables is restricted to this subsection.

Given (\ref{covSSC}) and (\ref{PPS})--(\ref{NWSSC}), the full components of both spin tensors, $S_{\mu\nu}$ and $\tilde S_{\mu\nu}$, are determined by $U^\mu$, $e_0{}^\mu$, and the \emph{covariant spin vector}
\begin{align}
S_{\mu}:=U^\nu (*S)_{\nu\mu}=U^\nu (*\tilde{S})_{\nu\mu},
\end{align}
satisfying $S_\mu U^\mu=0$, where $(*S)_{\nu\mu}=\frac{1}{2}\tensor{\epsilon}{_\nu_\mu^\alpha^\beta}S_{\alpha\beta}$ is the spin tensor's dual, and similarly for $(*\tilde{S})_{\nu\mu}$, with $\epsilon_{\mu\nu\alpha\beta}$ being the volume form.

\subsection{Multipole-curvature couplings}\label{sec:Lc}

So far, only the monopole and dipole couplings, arising from the first two terms in \eqref{poledipoleactiongeneral}, have been fixed.  Here we fix the couplings of the higher-order spin-induced multipoles to the spacetime curvature, in the $\mathcal L_c$ term in \eqref{poledipoleactiongeneral}, by considering all possible combinations of the relevant degrees of freedom which would contribute in a first post-Minkowskian approximation (but not yet making that approximation); -- this corresponds to keeping only terms linear in the Riemann tensor and its derivatives.

In constructing the curvature couplings, it is convenient to decompose the (vacuum) Riemann (or Weyl) tensor of the (external) gravitational field and all its symmetrized derivatives into their electric (even under parity) and magnetic (odd under parity) components with respect to the normalized velocity $U^\mu$, for $\ell\ge 2$, 
\begin{align}
\begin{aligned}
\mathcal{E}_{L}= & \ U_\alpha U_\beta \nabla^{\perp}_{(\mu_1}\ldots\nabla^{\perp}_{\mu_{\ell-2}}R^\alpha{}_{\mu_{\ell-1}}{}^\beta{}_{\mu_\ell)},\\
\mathcal{B}_{L}= & \ U_\alpha U_\beta \nabla^{\perp}_{(\mu_1}\ldots\nabla^{\perp}_{\mu_{\ell-2}}(*R)^\alpha{}_{\mu_{\ell-1}}{}^\beta{}_{\mu_\ell)},
\label{electricmagneticpart}
\end{aligned}
\end{align}
where $\nabla^{\perp}_\mu:=\mathcal P_\mu^\nu\nabla_\nu$, and $(*R)_{\alpha\beta\gamma\delta}=\frac{1}{2}\tensor{\epsilon}{_\alpha_\beta^\mu^\nu}R_{\mu\nu\gamma\delta}$ is the dual of the Riemann tensor. 

We can now build the linear-in-curvature couplings in $\mathcal L_c$ out of the $\mathcal E_{L}$'s and $\mathcal B_{L}$'s and the available point-particle degrees of freedom, namely, only the normalized tangent $U^\mu$ and the spin-vector $S_\mu$, recalling $S^\mu U_\mu=0$.  We require reparametrization invariance, which means that the tangent $u^\mu$ can enter only through its normalized version $U^\mu$, and invariance under internal rotations of the body-fixed tetrad $\Lambda_A{}^a$, which means that $\Lambda_A{}^a$ cannot enter explicitly.  We also require invariance under parity transformations, noting that $U^\mu$ and $\mathcal E_L$ are even while $S^\mu$ and $\mathcal B_L$ are odd. Then, all possible linear-in-curvature couplings are given by \cite{Levi:2015msa} (see also Refs.~\cite{Goldberger:2009qd, Ross:2012fc})
\begin{align}
\begin{aligned}
\mathcal{L}_{\text{c}}= -\sqrt{-u_\rho u^\rho}\bigg\{\sum_{\ell=1}^\infty \frac{(-1)^\ell}{(2\ell)!}\frac{C_{\mathcal{E},\ell}}{m^{2\ell-1}}S^{2L}\mathcal{E}_{2L} \\
-\sum_{\ell=1}^\infty \frac{(-1)^\ell}{(2\ell+1)!}\frac{C_{\mathcal{B},\ell}}{m^{2\ell}} S^{2L+1} \mathcal{B}_{2L+1}\bigg\},
\end{aligned}
\label{generalinteractionlagrangian}
\end{align}
where $C_{\mathcal{B},\ell}$ and $C_{\mathcal{E},\ell}$ are dimensionless constants (Wilson coefficients); the spacetime multi-indices are written out as $2L=\mu_1\ldots\mu_{2\ell}$ and $2L+1=\mu_1\ldots\mu_{2\ell+1}$.  Note, the monopole (no-spin) and dipole (spin-orbit) couplings, i.e.\ the $\ell=0$ terms, are not included here, since those are already provided outside of $\mathcal L_c$ in (\ref{poledipoleactiongeneral2}).

As written in \eqref{generalinteractionlagrangian}, $\mathcal{L}_c$ describes the higher-multipole couplings of an arbitrary body with spin-induced multipole moments. It is neglecting terms of quadratic or higher order in curvature, which would describe [adiabatic] tidal effects [assuming the bodies' internal dynamical time scales are much shorter than the orbital period, an assumption also implicit in (\ref{generalinteractionlagrangian})]. (For black holes, however, certain adiabatic tidal effects have been shown to vanish in four spacetime dimensions \cite{Damour:2009vw, Binnington:2009bb, Kol:2011vg, Gurlebeck:2015xpa, Pani:2015hfa}.) In order to specialize Eq.~\eqref{generalinteractionlagrangian} to the case where the multipoles match those of a spinning (Kerr) black hole, we set all the $C$ coefficients to unity: $C_{\mathcal{B},\ell}=C_{\mathcal{E},\ell}=1$ for all $\ell$. This can be justified retrospectively, e.g., by matching the binding energy of a BBH  with one spinning and one nonspinning black hole to the binding energy for a geodesic in the Kerr spacetime \cite{Bardeen:1972fi} at leading PN order.

\subsection{Linear approximation}\label{sec:linapp}

We now specialize to a first-post-Minkowskian (or linearized) approximation, writing the spacetime metric as
\begin{align}\label{g_eta_h}
g_{\mu\nu}=\eta_{\mu\nu}+h_{\mu\nu}+\order(G^2),
\end{align}
where $h_{\mu\nu}\sim\order(G)$ is the linear metric perturbation, using $G$ as a formal expansion parameter.  Ultimately, we aim for a leading-order post-\emph{Newtonian} (PN) approximation for the binary dynamics in the near zone of the source, which will be obtained in the following section by starting from the post-Minkowskian results discussed here and then reexpanding at leading orders in the orbital velocity.

When the effective point-particle action (for now, for just one black hole) is added to the Einstein-Hilbert action for the gravitational field, using the post-Minkowskian expansion (\ref{g_eta_h}) of the metric, we obtain a total effective action of the form
\begin{align}\label{Seff1}
S_{\text{eff}}[\mb{h},\mb{T}]= & \ S_{\text{G}}[\mathbf{h}]+S_{\text{kin}}[\mathbf{T}]+S_{\text{int}}[\mathbf{h},\mathbf{T}]+\order(G^2),
\end{align}
where $\mb{h}$ represents the degrees of freedom of the gravitational field, the metric perturbation $h_{\mu\nu}(x)$, and $\mb T=\{m,z(\lambda),\Lambda_A{}^a(\lambda),S^a(\lambda)\}$ represents the spinning point-particle degrees of freedom.
The term $S_{\text{G}}[\mathbf{h}]$ is the Einstein-Hilbert action at leading order in $\mb{h}$, which can be written as
\begin{align}
\mathcal{S}_G[\mathbf{h}] = - \frac{1}{64 \pi G} \int d^4x \, \partial_\rho h_{\mu\nu} P^{\mu\nu\alpha\beta} \partial^\rho h_{\alpha\beta},
\label{kinhaction}
\end{align}
while enforcing the harmonic gauge condition,
\begin{align}
\partial_\mu \bar{h}^{\mu\nu}=0,
\end{align}
where $\bar{h}^{\mu\nu}$ is the trace-reversed metric perturbation,
\begin{align}
\begin{aligned}
\bar{h}^{\mu\nu}&:=P^{\mu\nu\alpha\beta}h_{\alpha\beta},
\\
P^{\mu\nu\alpha\beta}& := \frac{1}{2} (\eta^{\mu\alpha} \eta^{\nu\beta} + \eta^{\nu\alpha} \eta^{\mu\beta} - \eta^{\mu\nu} \eta^{\alpha\beta} ).
\end{aligned}
\end{align}
The \emph{kinematic} term $S_{\text{kin}}[\mathbf{T}]$ and the \emph{interaction} term $S_{\text{int}}[\mathbf{h},\mathbf{T}]$ in \eqref{Seff1} are, respectively, the zeroth- and first-order terms in the expansion in $\mb h$ of the spinning point-particle action functional from Sec.~\ref{sec:effS},
\begin{align}\label{Sppsplit}
S_\text{p.p.}[\mb h,\mb T]=S_{\text{kin}}[\mathbf{T}]+S_{\text{int}}[\mathbf{h},\mathbf{T}]+\order(h^2).
\end{align}
We now proceed to find explicit forms for these terms while establishing appropriate conventions.

We choose to parameterize the worldline with coordinate time, $\lambda =t$, in an asymptotically Minkowskian coordinate system $x^\mu=(x^0,x^i)=(t,x^i)$. This implies $u^0=1$, as well as $u^i=v^i:=dx^i/dt$ for the 3-velocity. We define the usual Lorentz factor,
\begin{align}
\gamma=\frac{1}{\sqrt{1-v^2}}=\frac{1}{\sqrt{-u_\rho u^\rho}}+\order(h),
\end{align}
where, here and \emph{henceforth}, the Minkowski metric $\eta_{\alpha\beta}$ is used to raise and lower all spacetime indices; note also $v^2=v^iv^j\delta_{ij}$. The directional derivative along the particle's worldline is denoted with a dot
\begin{align}
\frac{d}{d\lambda}\rightarrow\frac{d}{dt}=:\dot{} \ .
\end{align}
We fix the freedom in the choice of the global tetrad $e_a{}^\mu$ by taking it to be the symmetric square root of the metric,
\begin{align}
e_a{}^\mu=\delta_a^\nu\left(\delta^{\mu}_{\nu}-\frac{1}{2}h^{\mu}{}_{\nu}\right)+\order(h^2).
\end{align}

With these conventions, taking the $\order(h^0)$ part of $S_\text{p.p.}$ from \eqref{poledipoleactiongeneral2} (with tildes removed) to obtain $S_\text{kin}$ in \eqref{Sppsplit}, noting that $\Lambda_0{}^\mu=\delta_0^\mu$ as follows from \eqref{tetradconstraint} and \eqref{boosttetrad}--\eqref{boost}, we find
\begin{align}
S_{\text{kin}}[\mathbf{T}]= \int dt \bigg\{-\frac{m}{\gamma} +\frac{1}{2}S_{ij}\tensor{\Lambda}{_K^i}\tensor{\dot{\Lambda}}{^K^j}
+\frac{1}{2}S_{ij}v^i\dot{v}^j \bigg\},
\label{kinTaction}
\end{align}
with $K=1,2,3$.  

For the $\order(h^1)$ interaction part of $S_\text{p.p.}$ in \eqref{Sppsplit}, we write $S_\text{int}=\int dt \,(\mathcal{L}^{\text{pole-}}_{\text{dipole}}+\mathcal{L}_{\text{c}})$, and the pole-dipole terms, arising from the expansions of the first and fourth terms in \eqref{poledipoleactiongeneral2}, are found to be
\begin{align}
\mathcal{L}^{\text{pole-}}_{\text{dipole}}= & \  \frac{m}{2 \gamma}U^\mu U^\nu\bigg[ h_{\mu\nu}+\tensor{\epsilon}{_\nu_\rho^\alpha^\beta} a^\rho\partial_\alpha h_{\beta\mu}\bigg]+\order(\dot h),
\end{align}
where we use the mass-rescaled spin vector
\begin{align}
a^\mu:=\frac{S^\mu}{m}.
\end{align}
Notice, we neglect here time derivatives of $h_{\mu\nu}$; via integration-by-parts, noting $\dot{U}^\mu=\order(h)$ and $\dot{a}^\mu=\order(h)$, these become $O(h^2)$ terms and total time derivative terms. (Neglecting the total time derivatives corresponds to an implicit redefinition of variables \cite{Damour:1990jh}.)

Finally, for the $\order(h)$ part of the higher-multipole couplings in $\mathcal L_c$ \eqref{generalinteractionlagrangian}, we can use the expansions
\begin{align}
\begin{aligned}
\mathcal{E}_{\alpha\beta}= & \ -\frac{1}{2}U^\mu U^\nu\partial_\alpha\partial_\beta h_{\mu\nu}+\order(\dot{h},h^2),\\
\mathcal{B}_{\alpha\beta}= & \ \frac{1}{2}U^\mu U^\nu \tensor{\epsilon}{_\nu_{(\beta}^\gamma^\tau}\partial_{\alpha)}\partial_\gamma h_{\tau\mu}+\order (\dot{h},h^2)
\end{aligned}
\end{align}
of the curvature tensors.  We then find that it is natural to combine the contributions from $\mathcal{L}^{\text{pole-}}_{\text{dipole}}$ and $\mathcal L_c$ by extending the sums in  \eqref{generalinteractionlagrangian} to $\ell=0$.  Having fixed $C_{\mathcal{B},\ell}=C_{\mathcal{E},\ell}=1$ to ensure matching with the Kerr spacetime, as outlined above, the full interaction term is expressed as 
\begin{align}
\begin{aligned}
S_{\text{int}}[\mathbf{h},\mathbf{T}]= & \int dt \ \bigg\{\sum_{\ell=0}^\infty \frac{m U^{\mu} U^{\nu}}{2 \gamma \, \ell!} \text{Re}\bigg[i^\ell a^L \partial_L h_{\mu\nu} \\
& + i^{\ell -1} a^\rho \tensor{\epsilon}{_\nu_\rho^\alpha^\beta} a^{L-1} \partial_\alpha \partial_{L-1} h_{\beta\mu}\bigg]\bigg\},
\label{BHfullaction}
\end{aligned}
\end{align}
which matches a result in \cite{Vines:2016qwa}.  Having completed the relevant post-Minkowskian expansions, we henceforth set $G=1$.

\section{Conservative Dynamics} \label{sectionconservativedynamics}

The building blocks from the previous section are now combined to give an effective description of a BBH at leading post-Newtonian order but to all orders in spin. The underlying field equations are derived from the full effective BBH action
\begin{align}
\begin{aligned}
S_{\text{eff}}^{\text{BBH}}&[\mathbf{h},\mathbf{T}_1,\mathbf{T}_2]= S_G[\mathbf{h}] \\
& +\{S_{\text{kin}}[\mathbf{T}_1] +S_{\text{int}}[\mathbf{h},\mathbf{T}_1]+(1\leftrightarrow 2)\},
\label{BBHaction}
\end{aligned}
\end{align}
with two copies of the kinetic and interaction terms as discussed above for each of the black holes, 1 and 2.
A slow-motion approximation is achieved by expansion in the orbital velocity $v$ up to linear order. Our near-zone (NZ) solution $h_{\text{NZ}}^{\mu\nu}$, obtained in Sec.~\ref{fieldequationssection} in the no-retardation limit, agrees with results presented in \cite{Vines:2017hyw,Harte:2016vwo}. We follow the Fokker-action approach used in \cite{Bernard:2015njp} to derive the equations of motion (EOM) of the orbital parameters in Sec.~\ref{eomsection}. A Hamiltonian encoding equivalent EOM has been obtained using effective field theoretic tools in \cite{Vines:2016qwa}. We then put the EOM into explicit form. Finally, in Sec.~\ref{orbitalparametersection}, we solve these EOM, under the above assumption of circular spin-aligned motion, and obtain the conserved energy and conserved orbital angular momentum of the BBH to all orders in the spins and at the leading PN orders.

\subsection{Field equations and near-zone solution} \label{fieldequationssection}

In the full effective BBH action, the metric perturbation was considered at leading post-Minkowskian order. The linearized field equations, in harmonic coordinates, are obtained by varying \eqref{BBHaction} with respect to the fields $h_{\mu\nu}$. Integrating by parts while dropping vanishing boundary terms yield the field equations 
\begin{align}
\begin{aligned}
\square \bar{h}^{\mu\nu}_A= & \ -16\pi T^{\mu\nu}_A, \\
T^{\mu\nu}_A= & \ m_A\gamma_A \sum_{\ell=0}^{\infty}\frac{(-1)^\ell}{\ell!}\text{Re}\Big[i^\ell u^{\mu}_Au^{\nu}_Aa^L_A\partial_L\delta_A \\
& +i^{\ell-1}u^{\sigma}_Aa^{\rho}_A\tensor{\epsilon}{_\sigma_\rho^\alpha^{(\mu}}u_A^{\nu)}a^{L-1}_A\partial_{\alpha L-1}\delta_A\Big],
\label{fieldequations}
\end{aligned}
\end{align}
for black holes $A=1,2$, where we split $\bar{h}^{\mu\nu}=\bar{h}^{\mu\nu}_1+\bar{h}^{\mu\nu}_2$.  We have used $\int dt=\int d^4x \,\delta_A$, with \mbox{$\delta_A:=\delta[\boldsymbol{x}-\boldsymbol{z}_A(t)]$} in \eqref{BHfullaction}.

The general solution to the inhomogeneous wave equation is well-known. The retarded inverse d'Alembertian integral operator, defined by $\bar{h}^{\mu\nu}=-16 \pi \square^{-1}_{\text{ret}}T^{\mu\nu}$, reduces as $(\square^{-1}_{\text{ret}}T^{\mu\nu})(\boldsymbol{x},t)\rightarrow (\Delta^{-1}T^{\mu\nu})(\boldsymbol{x},t)$ at leading PN order in the NZ, where \cite{Blanchet:2013haa}
\begin{align}
(\Delta^{-1}T^{\mu\nu})(\boldsymbol{x},t)= -\frac{1}{4\pi} \int d^3x' \frac{T^{\mu\nu}(\boldsymbol{x}',t)}{|\boldsymbol{x}-\boldsymbol{x}'|}.
\end{align}
Retardation effects would contribute only at next-to-leading PN orders. Applying $\Delta^{-1}$ to the effective energy-momentum tensor in \eqref{fieldequations}, the NZ linearized gravitational field of the $A$th black hole in the binary is explicitly given by
\begin{align}
\begin{aligned}
\bar{h}^{\mu\nu}_{\text{NZ},A}(t,\boldsymbol{x})= & \ 4m_A\gamma_A\sum_{\ell=0}^{\infty}\frac{(-1)^\ell}{\ell!}\text{Re}\Big[i^\ell u^{\mu}_Au^{\nu}_Aa^L_A\partial_Lr^{-1}_A\\
&+i^{\ell-1}u^{\sigma}_Aa^{\rho}_A\tensor{\epsilon}{_\sigma_\rho^\alpha^{(\mu}}u^{\nu)}_Aa^{L-1}_A\partial_{\alpha L-1}r^{-1}_A\Big],
\label{NZnonlinsolution}
\end{aligned}
\end{align}
with $r_A:=|\boldsymbol{x}-\boldsymbol{z}_A(t)|$. This solution has been obtained in the linear post-Minkowskian approximation, but it still contains nonlinear-in-velocity contributions at each order in spin.

The leading-order slow-motion approximation is achieved by truncating the NZ solutions \eqref{NZnonlinsolution} after linear-in-$v$ terms. This yields a leading PN expansion at each order in spin. Carefully excluding the higher-order in velocity terms (e.g., noticing that $\gamma=1+\order(v^2)$ and $a^0=\order(v)$), the trace $h_{\text{NZ},A}$ of the solution $h^{\mu\nu}_{\text{NZ},A}=\tensor{P}{^\mu^\nu_\alpha_\beta}\bar{h}^{\alpha\beta}_{\text{NZ},A}$ is given by
\begin{align}
\begin{aligned}
h_{\text{NZ},A}= & \ \bar{h}^{00}_{\text{NZ},A}-\bar{h}^{ij}_{\text{NZ},A}\delta_{ij} \\
= & \ 4m_A \mathcal{D}_C[\boldsymbol{a}_A] r^{-1}_A+\mathcal{O}(v^2).
\label{NZtrace}
\end{aligned}
\end{align}
Here, we introduced the notation
\begin{align}
\begin{aligned}
\mathcal{D}_C[\boldsymbol{a}]r^{-1}_A:=& \ \sum_{\ell =0}^{\infty} \frac{(-1)^\ell}{(2\ell )!}a^{2L}\partial_{2L}r^{-1}_A\\
= & \ \cosh(\boldsymbol{a}\times \nabla)r^{-1}_A,
\label{diffopC}
\end{aligned}
\end{align}
noting that $r_A\neq 0$ and define for later convenience
\begin{align}
\begin{aligned}
\mathcal{D}^i_S[\boldsymbol{a}]r^{-1}_A:=& \ -a^j\tensor{\epsilon}{_j^i^k}\sum_{\ell =0}^{\infty} \frac{(-1)^\ell}{(2\ell +1)!}a^{2L}\partial_{2Lk}r^{-1}_A\\
= & \ [\sinh(\boldsymbol{a}\times \nabla)]^ir^{-1}_A.
\label{diffopS}
\end{aligned}
\end{align}
Here, $\nabla=\partial/\partial \boldsymbol{x}$, and the index $i$ in \eqref{diffopS} labels the components of the operator with respect to the chosen basis (i.e., can be raised and lowered with $\delta_{ij}$). Additionally, both operators obey the usual hyperbolic trigonometric identities.
\begin{widetext}
\noindent The NZ gravitational fields $h^{\mu\nu}_{\text{NZ},A}$ of a spinning black hole at 1PM and to linear order in velocities are
\begin{subequations}
\begin{align}
h^{00}_{\text{NZ},A}= & - 2\phi_A +\mathcal{O}(v^2), 
& \phi_A:= & \ \Big\{-\mathcal{D}_C[\boldsymbol{a}_A]+2v_i^{(A)}\mathcal{D}^i_S[\boldsymbol{a}_A]\Big\}\frac{m_A}{r_A},
\label{solution2a} \\
h^{0j}_{\text{NZ},A}= & \ A^j_A +\mathcal{O}(v^2), 
& A^i_A:= & \ \Big\{4v^i_A\mathcal{D}_C[\boldsymbol{a}_A]-2\mathcal{D}^i_S[\boldsymbol{a}_A] \Big\}\frac{m_A}{r_A},\\
h^{ij}_{\text{NZ},A}= & \ \frac{1}{2}h_{\text{NZ},A}\delta^{ij}+\sigma^{ij}_A +\mathcal{O}(v^2), 
& \sigma^{ij}_A:= & \ -4v^{(j}_A\mathcal{D}^{i)}_S[\boldsymbol{a}_A]\frac{m_A}{r_A}.
\label{solution2c}
\end{align}
\label{NZsolution}
\end{subequations}
\end{widetext}
The complete leading PN order NZ solution of the BBH
\begin{align}
h_{\mu\nu}^{\text{NZ},\text{BBH}}=h_{\mu\nu}^{\text{NZ},1}+h_{\mu\nu}^{\text{NZ},2}+\order(v^2)
\label{NZfield}
\end{align}
in this linear approach, is constructed by superposing the gravitational fields of both black holes.

\subsection{EOM for the separation vector} \label{eomsection}

In the following, the focus lies on the derivation of the set of EOM describing the binary's separation vector $\bs{r}:=\boldsymbol{z}_1-\boldsymbol{z}_2$, with $r:=|\bs{r}|$. From this, the spin corrections to the Newtonian orbital parameter $\omega$, the angular velocity, are acquired. These equations for $\bs{r}$ result from the kinematic behavior of the black holes in the time-dependent near-zone field \eqref{NZfield}. In the framework of the Fokker-action approach, this behavior is encoded in
\begin{align}
\begin{aligned}
S_{\text{eff}}^{\text{EOM}}[\mathbf{h}^{\text{NZ},1},\mathbf{h}^{\text{NZ},2},\mathbf{T}_1,\mathbf{T}_2]= & \ \frac{1}{2} S_{\text{int}}[\mathbf{h}^{\text{NZ},2},\mathbf{T}_1]\\
& +S_{\text{kin}}[\mathbf{T}_1]+(1\leftrightarrow 2),
\end{aligned}
\label{eomeffaction}
\end{align}
where we again made use of the functionals \eqref{kinTaction} and \eqref{BHfullaction}. Note, we utilized $S_G[\mathbf{h}^{\text{NZ},1}+\mathbf{h}^{\text{NZ},2}]=-1/2 S_{\text{int}}[\mathbf{h}^{\text{NZ},1},\mathbf{T}_2]+(1\leftrightarrow 2)$, plus a total derivative of (divergent) $h_1^2$ and $h_2^2$ terms which do not influence the EOM. The equation relating the angular frequency $\omega$ to the radius $r$ will be obtained from the EOM for $\bs{r}$, restricted to circular spin-aligned motion. 

One of the interaction terms in \eqref{eomeffaction} is given by
\begin{align}
\begin{aligned}
S_{\text{int}}[\mathbf{h}^{\text{NZ},2},\mathbf{T}_1]=\frac{m_1\gamma_1}{2}\int dt\sum_{\ell =0}^{\infty}\frac{1}{\ell!}\text{Re}\Big[i^\ell u^{\mu}_1u^{\nu}_1a^L_1\partial_L 
\\
+i^{\ell -1}u^{\sigma}_1a^{\rho}_1\tensor{\epsilon}{_\sigma_\rho^\alpha^{\mu}}u^{\nu}_1a^{L-1}_1\partial_{\alpha L-1}\Big]h_{\mu\nu}^{\text{NZ},2}\bigg|_{\boldsymbol{z}_1}.
\end{aligned}
\end{align}
By redefining $\nabla:=\partial/\partial \boldsymbol{z}_1$ and noting that the differential operators $\mathcal{D}_C[\boldsymbol{a}]$ and $\mathcal{D}^i_S[\boldsymbol{a}]$ in \eqref{diffopC} and \eqref{diffopS} change accordingly, the interaction terms can be combined into
\begin{align}
\begin{aligned}
& S_{\text{int}}[\mathbf{h}^{\text{NZ},2},\mathbf{T}_1]+S_{\text{int}}[\mathbf{h}^{\text{NZ},1},\mathbf{T}_2]= \\
& \int dt \ \bigg\{\bigg[ 2\mathcal{D}_C[\boldsymbol{a}_0]+ 4v_i\mathcal{D}^i_S[\boldsymbol{a}_0]\bigg]\frac{m_1m_2}{r} +\order(v^2) \bigg\}.
\label{Sint}
\end{aligned}
\end{align}
Hyperbolic trigonometric identities have been used to fuse the operators, and we have defined 
\begin{align}
\begin{aligned}
\boldsymbol{a}_0:=\boldsymbol{a}_1+\boldsymbol{a}_2, \qquad \boldsymbol{v}:=\boldsymbol{v}_1-\boldsymbol{v}_2.
\end{aligned}
\end{align} 

The kinetic terms follow directly from \eqref{kinTaction}. With $S_{ij}=\epsilon_{ijk}S^k$ and ${\Omega}_k:=\frac{1}{2}\epsilon_{ijk}\Omega^{ij}$, the sum of the kinetic terms in \eqref{eomeffaction} reads
\begin{align}
\begin{aligned}
S_{\text{kin}} & [\mathbf{T}_1] +S_{\text{kin}}[\mathbf{T}_2]=\int dt \ \bigg\{-m_1 +\frac{m_1}{2}v_1^2 \\
&  +\frac{1}{2}\boldsymbol{S}_1\cdot(\boldsymbol{v}_1\times
\dot{\boldsymbol{v}}_1) +\bs{S}_1\cdot {\bs{\Omega}}_1 +(1\leftrightarrow 2)\bigg\}.
\end{aligned}
\label{Skin}
\end{align}
Then, the effective action \eqref{eomeffaction}, together with \eqref{Sint} and \eqref{Skin}, provides the full conservative dynamics of the binary, encoded in the Lagrangian
\begin{align}
\begin{aligned}
\mathcal L= & \ \bigg[-m_1 +\frac{m_1}{2}v_1^2 +\frac{1}{2}\boldsymbol{S}_1\cdot(\boldsymbol{v}_1\times
\dot{\boldsymbol{v}}_1) +\bs{S}_1\cdot {\bs{\Omega}}_1 \\ 
 & + (1\leftrightarrow 2) \bigg] +  \bigg[ \mathcal{D}_C[\boldsymbol{a}_0]+ 2v_i\mathcal{D}^i_S[\boldsymbol{a}_0] \bigg]\frac{m_1 m_2}{r}.
\end{aligned}
\label{Leff}
\end{align}

\begin{figure}[t]
\centering
\includegraphics[scale=1.5]{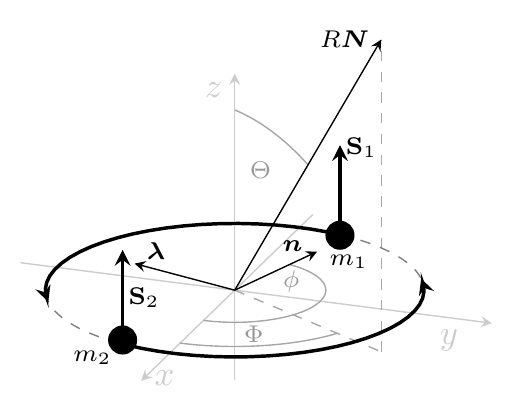}
\caption{The BBH configuration is illustrated for the case of $m_1=m_2$ and $\boldsymbol{S}_1=\boldsymbol{S}_2$. The coordinates are chosen such that the orbital plane coincides with the $(x,y)$-plane: $\hat{z}\equiv \bs{\ell}$. Finally, $\bs{N}$, introduced together with $R$ in \eqref{Nvector}, is pointing radially outwards to the field point, parameterized by $\Theta$ and $\Phi$ in the usual way.}
\label{plot}
\end{figure}

At this stage, it is convenient to specialize the coordinate system on the spatial slices, by choosing the coordinate origin to be the center of mass of the BBH. This amounts to setting the Noether charge associated with the boost symmetry of \eqref{Leff} to zero (see Appendix~\ref{app:Noe}), resulting in
\begin{align}
\begin{aligned}
\begin{split}
\boldsymbol{z}_1=\frac{m_2}{M}\bs{r}-\bs{b}, \qquad \boldsymbol{z}_2=-\frac{m_1}{M}\bs{r}-\bs{b}, 
\end{split} \\
\begin{split}
\quad \bs{b}:=\frac{1}{M}(\bs{v}_1\times\boldsymbol{S}_1+\bs{v}_2\times\boldsymbol{S}_2),
\end{split}
\label{COMframe}
\end{aligned}
\end{align}
where we define the system's total mass $M$ and for future convenience the reduced mass $\mu$,
\begin{align}
M:=m_1+m_2,\qquad\mu:=\frac{m_1m_2}{M}.
\end{align}
Similar relations are found for the velocities $\bs{v}_A=\dot{\boldsymbol{z}}_A$. The unit vectors 
\begin{align}
\bs{n}:=\frac{\bs{r}}{r}, & & \boldsymbol{\lambda}:=\frac{\bs{v}}{v},
\end{align}
with $v:=|\bs{v}|$, span the orbital plane. Finally, a third vector $\bs{\ell}$ is constructed to be orthonormal with the orbital plane. Then, the three vectors $\{\bs{n},\boldsymbol{\lambda},\boldsymbol{\ell}\}$ pose a positively oriented triad, $\bs{n}\times\boldsymbol{\lambda}= \boldsymbol{\ell}$. In \figurename{ \ref{plot}}, the BBH configuration in the center-of-mass frame is depicted for the case of equal masses and spin vectors.

Using \eqref{COMframe} to specialize the Lagrangian \eqref{Leff} to the center-of-mass frame, the translational degrees of freedom are reduced from the two positions $\boldsymbol{z}_{1}(t)$ and $\boldsymbol{z}_{2}(t)$ to the single separation vector $\boldsymbol{r}(t)=\boldsymbol{z}_{1}-\boldsymbol{z}_{2}$.  Working consistently within the leading-PN-order expansion (where it is sufficient to use the \emph{leading-order} relations $\boldsymbol{v}_1=m_2\boldsymbol{r}/M$ and $\boldsymbol{v}_2=-m_1\boldsymbol{r}/M$ and their time derivatives and $\dot{\boldsymbol{S}}_{1,2}=0$), we find the following final reduced form of the Fokker-type (generalized) Lagrangian for the BBH,
\begin{align}
\begin{aligned}
\mathcal L&= \frac{\mu v^2}{2} + \frac{\mu}{2}\boldsymbol{\sigma}^*\cdot(\boldsymbol{v}\times\dot{\boldsymbol{v}}) +\bs{S}_1\cdot {\bs{\Omega}}_1 + \bs{S}_2\cdot {\bs{\Omega}}_2
\\
 &\quad+  \bigg[ \mathcal{D}_C[\boldsymbol{a}_0]+ 2v_i\mathcal{D}^i_S[\boldsymbol{a}_0] \bigg]\frac{m_1 m_2}{r},
\end{aligned}
\label{mcLfinal}
\end{align}
using \eqref{diffopC}--\eqref{diffopS} with $\partial_i=\partial/\partial r^i$, recalling $\boldsymbol{a}_A=\boldsymbol{S}_A/m_A$ and $\boldsymbol{a}_0=\boldsymbol{a}_1+\boldsymbol{a}_2$, and defining
\begin{align}
\bs{\sigma}^*:=\frac{m_2\boldsymbol{a}_1+m_1\boldsymbol{a}_2}{M}.
\end{align}
Applying the generalized Euler-Lagrange equation,
\begin{align}
0= \frac{\partial \mathcal L}{\partial \bs{r}}-\frac{d}{dt} \frac{\partial \mathcal L}{\partial \bs{v}}+\frac{d^2}{dt^2}\frac{\partial \mathcal L}{\partial \dot{\bs{v}}},
\label{ELr}
\end{align}
with appropriate reduction of order,\footnote{By ``reduction of order'', we mean the following.  In solving for the acceleration $\dot{\bs v}=\ddot{\bs r}$ in terms of $\bs r$ and $\bs v$, working consistently within the PN expansion, we can substitute the zeroth-order [$O(v^0)$] solution for the acceleration where it appears in first-order [$O(v^1)$] terms. Note that it is important that we do not make this replacement until after we have varied the action to obtain the EOM. If we were to make such a replacement of $\dot{\bs v}$ in the Lagrangian \eqref{mcLfinal}, we would obtain an equivalent form of the action but one in which the variables have been implicitly redefined; specifically, the separation $\bs r=\bs z_1-\bs z_2$ (between the covariant-SSC worldlines which we use here) would become the separation $\tilde{\bs r}=\tilde{\bs z}_1-\tilde{\bs z}_2$ between the canonical-SSC worldlines as discussed in footnote \ref{footNWSSC}.  The two are related at leading order by
\begin{align}
\tilde{\bs r}=\bs r+\frac{1}{2}\bs v\times\bs\sigma^*.
\end{align}
}
we find the EOM for the separation vector $\bs{r}$:
\begin{align}
\begin{aligned}
\dot v^j=\ddot{r}^j= \ \bigg\{ & \partial^j \mathcal{D}_C[\boldsymbol{a}_0] + v^i\sigma^{*k}\tensor{\epsilon}{_k_l^j}\partial_i\partial^l \mathcal{D}_C[\boldsymbol{a}_0]\\
&- 2\Big[\delta^j_iv^k\partial_k-\partial^j v_i\Big]\mathcal{D}^i_S[\boldsymbol{a}_0]\bigg\}\frac{M}{r}.
\end{aligned}
\label{zeom}
\end{align}
The action of the differential operators $\mathcal{D}_C[\boldsymbol{a}_0]$ and $\mathcal{D}^i_S[\boldsymbol{a}_0]$ on $r^{-1}$ is presented in detail in Appendix~\ref{appendixB}. Utilizing the results shown there, derived under the assumption of circular and spin-aligned motion, the implicit EOM \eqref{zeom} simplify to the explicit form
\begin{align}
\begin{aligned}
\ddot{\bs{r}}= & -\bs{r} \bigg[\frac{M}{(r^2-a^2_0)^{3/2}}-\frac{vM(\sigma^*+2a_0)}{r(r^2-a_0^2)^{3/2}}\bigg],
\label{zeomsimple}
\end{aligned}
\end{align}
where $a_0:=\boldsymbol{\ell}\cdot \boldsymbol{a}_0$ and $\sigma^*:=\boldsymbol{\ell}\cdot \boldsymbol{\sigma}^*$.

\subsection{Orbital parameter and conserved energy} \label{orbitalparametersection}

\begin{table*}[t]
\begin{ruledtabular}
\centering
\begin{tabular}{cc|ccccccccc}
&  & $S^0$ & $S^1$ & $S^2$ & $S^3$ & $S^4$ &  $\dots$ & $S^{p}$\footnotemark[1] & $S^{p+1}$\footnotemark[1] & $\dots$ \\ \hline \hline
\multirow{2}{*}{LO} & 0PN & $A_0\cdot 1$ &  & $A_2\cdot x^2 \chi^2$ &  & $A_4\cdot x^4\chi^4$ & $\dots$ & $A_p\cdot x^{p}\chi^p$ &  & $\dots$ \\ \cline{2-11}
& 0.5PN &  & $A_1\cdot x^{3/2}\chi$ &  & $A_3\cdot x^{7/2}\chi^3$ &  & $\dots$ &  & $A_{p+1}\cdot x^{p+3/2}\chi^{p+1}$ & $\dots$ \\ \hline 
\multirow{2}{*}{NLO} & 1PN & $B_0\cdot x$ &  & $B_2\cdot x^3 \chi^2$ &  & $B_4\cdot x^5\chi^4$ & $\dots$ & $B_{p}\cdot x^{p+1}\chi^{p}$ &  & $\dots$ \\ \cline{2-11}
& 1.5PN &  & $B_1\cdot x^{5/2}\chi$ &  & $B_3\cdot x^{9/2}\chi^3$ &  & $\dots$ &  & $B_{p+1}\cdot x^{p+5/2}\chi^{p+1}$ & $\dots$ \\ \hline 
$\vdots$ & $\vdots$ & $\vdots$ & $\vdots$ & $\vdots$ & $\vdots$ & $\vdots$ &  & $\vdots$ & $\vdots$ & $\ddots$ \\
\end{tabular} 
\footnotetext[1]{Here $p$ is even.}
\caption{The results presented in this paper are expansions \textit{separately} in the traditional post-Newtonian parameter $\epsilon_{\text{PN}}\sim v^2\sim m/r\sim x$ and the spin expansion parameter $\epsilon_{\text{spin}}\sim a/r\sim x \chi$ (In this table, $\chi$ serves as book keeping parameter for the spin expansion). Different rows correspond to different powers of $\epsilon_{\text{PN}}$ (LO stands for leading-order, and NLO for next-to-leading order) and different columns to different powers of $\epsilon_{\text{spin}}$.  Let $p$ be even. Then there is an absolute leading order term of the expansion in $x\sim \epsilon_{\text{PN}}$, given by $A_p\cdot\epsilon_{\text{spin}}^p=A_p\cdot x^{p}\chi^p$, with coefficient $A_p=A_p(m_1,m_2,\boldsymbol{a}_1,\boldsymbol{a}_2)$. This is the leading post-Newtonian term at that order in spin. Similary, at ($p$+1)th order in spin, the absolute leading order term, in the expansion of $x\sim\epsilon_{\text{PN}}$, is $A_{p+1}\cdot \sqrt{\epsilon_{\text{PN}}} \epsilon_{\text{spin}}^{p+1}=A_{p+1}\cdot x^{p+3/2}\chi^{p+1}$. In this work, we only focus on the leading PN order, i.e., $B_p\equiv 0$ for all $p$, which we denote by LO-S$^\infty$. Every expression (e.g., conserved energy, total energy flux, etc.) can be written as a sum over $p$ of all LO terms with different coefficients $A_p$ (up to a multiplicative function depending on masses and $x$).} \label{PNorder}
\end{ruledtabular}
\end{table*}

The EOM \eqref{zeomsimple} allow a wide variety of different solutions for the separation vector $\bs{r}$. However, in this paper we only focus on the special case of circular motion with aligned spins. In this case, the spins are constant vectors (as can be verified from their EOM, not presented here), and the acceleration $\ddot{\bs{r}}$ is directly proportional to $\bs{r}$, with the squared angular velocity serving as proportionality constant:
\begin{align}
\ddot{\bs{r}}=-\omega^2\bs{r}.
\label{definitionomega}
\end{align}
Comparing \eqref{zeomsimple} and \eqref{definitionomega}, we find
\begin{align}
\omega^2= & \ \frac{M}{r}\frac{r-v(2a_0+\sigma^*)}{(r^2-a_0^2)^{3/2}},
\end{align}
for the radius-frequency relationship.  For convenience, we introduce the standard PN expansion parameter
\begin{align}\label{xomega}
x=(\omega M)^{2/3}.
\end{align}
Then, the separation $r$ of the black holes at leading post-Newtonian order and to all orders in spin, denoted by LO-S$^\infty$, is split into even- and odd-in-spin parts 
\begin{subequations}\label{rofx}
\begin{align}
r_{\text{LO-S}^{\infty}}(x)=r_{\text{even}}(x)+r_{\text{odd}}(x),
\end{align}
where
\begin{align}
\begin{aligned}
r_{\text{even}}(x):= & \ \sqrt{\frac{M^2}{x^2}+a_0^2},\\ 
r_{\text{odd}}(x):= & \ -r_{\text{even}}(x)\frac{x^{3/2}M}{3}\frac{\sigma^*+2a_0}{M^2+x^2a_0^2}.
\end{aligned}
\end{align}
\end{subequations}
Expanding this result yields a leading PN spin expansion of the form described in \tablename{ \ref{PNorder}}.

The conserved total energy $E$ and total angular momentum $\bs J$ of the BBH system can be obtained as the Noether charges associated with the time-translation and rotation symmetries of the effective action, as discussed in Appendix \ref{app:Noe}. For circular orbits and aligned spins, 
we find
\begin{align}
E=\mu\left(\frac{1}{2}r^2\omega^2+r^2\omega^3\sigma^*-\frac{M}{\sqrt{r^2-a_0^2}}\right),
\label{conservedenergy}
\end{align}
and
\begin{align}
J=\mu \left(\omega r^2+\frac{3}{2}\omega^2r^2\sigma^*-\frac{2M a_0}{\sqrt{r^2-a_0^2}} \right)+S_1+S_2,
\label{conservedanglmoment}
\end{align}
with $J:=\bs{J}\cdot\bs{\ell}$ and similarly for $S_{1,2}$.
Finally, using $r(x)$ from \eqref{rofx}, we find $E$ and $J$ as functions of the frequency parameter $x$ from \eqref{xomega}, given to all orders in spin at leading post-Newtonian order by
\begin{align}
\begin{aligned}
E_{\text{LO-S}^{\infty}}(x)= -\frac{\mu x}{2}\Big\{ & 1+\frac{x^{3/2}}{3M}(7 a_0+\delta a_-)-\frac{x^2 a_0^2}{M^2} \\
& -\frac{x^{7/2}a_0^2}{M^3}(a_0-\delta a_-)\Big\},
\end{aligned}
\end{align}
and
\begin{align}
\begin{aligned}
J_{\text{LO-S}^\infty}(x)=\mu x^{-1/2}\bigg\{M-\frac{5}{12}x^{3/2}(7a_0+\delta a_-) \\
+\frac{x^2 a_0^2}{M} +\frac{3x^{7/2}}{4M^2}a_0^2(a_0-\delta a_-) \bigg\}+S_1+S_2,
\end{aligned}
\end{align}
where we define the symmetric mass ratio $\nu$, the antisymmetric mass ratio $\delta$, and the antisymmetric spin combination $\boldsymbol{a}_-$ with $a_-:=\bs{\ell}\cdot\boldsymbol{a}_-$,
\begin{align}
\nu:=\frac{\mu}{M},
\qquad
\delta:=\frac{m_1-m_2}{M},
\qquad
\boldsymbol{a}_-:=\boldsymbol{a}_1-\boldsymbol{a}_2.
\end{align}

Remarkably, the spin-expansions of the binding energy $E$ and orbital angular momentum $L:=J-S_1-S_2$ terminate after cubic-in-spin contributions.  Thus, our all-orders-in-spin result for $E$ coincides with the cubic-in-spin result of \cite{Marsat:2014xea} for the case of aligned spins; we have extended the validity of the expression presented there to all orders in spin.  Note that the truncation at cubic order in spin hinges on the use of $x$ (or $\omega$) as the variable.  Compare also to the $E(x)$ result for neutron stars to quartic order in spin \cite{Levi:2016ofk}, where the constants $C_{\mathcal{E},\ell}$ and $C_{\mathcal{B},\ell}$ are not unity as for black holes, and the fourth-order-in-spin coefficients do not vanish.

The justification for the choice of coefficients in \eqref{generalinteractionlagrangian} can now be presented. We define the dimensionless Kerr spin-parameter $\bs{\chi}_A=m_A\boldsymbol{a}_A$ of the individual black holes in the considered BBH. Then, in the limit of $m_2/m_1\rightarrow 0$, the conserved energy $E_{\text{LO-S}^\infty}$ reduces to the binding energy associated with geodesic motion in Kerr spacetime \cite{Bardeen:1972fi} (characterized by spin-parameter $m\chi$), under the identification $\bs{\chi}_1\cdot \bs{\ell}\rightarrow \chi$. A small deviation of the coefficients $C_{\mathcal{E},\ell}$ and $C_{\mathcal{B},\ell}$ from unity would have led to additional terms in the test-body limit of $E_{\text{LO-S}^\infty}$. The resulting binding energy would not have matched the one for geodesic motion in the Kerr solution. Thus, $C_{\mathcal{E},\ell}=C_{\mathcal{B},\ell}=1$ is the unique choice to approximate the Kerr solution at 1PM order.

\section{Far zone modes and energy flux} \label{sectionfluxandmodes}

In this section, the far zone dynamics of the binary black hole is analyzed. Again, this is done in the leading post-Newtonian approximation scheme employed before, where all spin-induced multipole moments are considered. The gravitational wave modes and the total energy flux emitted by the binary are determined at future null infinity in Sec.~\ref{fluxsection} and Sec.~\ref{modessection}, respectively. To do so, we construct a set of source multipole moments, of the complete BBH, in Sec.~\ref{sourcemomentssection}. Our results for these source moments agree with those presented in \cite{Marsat:2014xea} (before inserting the solution for the orbital separation $r$). 

\subsection{Source multipole moments} \label{sourcemomentssection}

We consider the mass and current type source multipole moments, $\mathcal{I}_L(\tilde{t})$ and $\mathcal{J}_L(\tilde{t})$, respectively, at leading post-Newtonian order. Here, we introduced the Euclidean distance $R$ from the above defined center of mass of the system to a far zone spacetime point, which enables us to include the retardation by $\tilde{t}:=t-R$. Notice, differentiation with respect to $\tilde{t}$ and $t$ is equivalent and will be denoted by a dot as before.

The complete energy-momentum distribution of the BBH, the sum of the individual contributions \eqref{fieldequations}, is
\begin{align}
\begin{aligned}
T^{00}= & \left\{\mathcal{D}_C[\boldsymbol{a}_1]-v^{(1)}_i\mathcal{D}^i_S[\boldsymbol{a}_1]\right\}m_1\delta_{1}+ (1\leftrightarrow 2) +\mathcal{O}(v^2), \\
T^{0j}= & \ \frac{1}{2}\left\{ 2 v_1^j\mathcal{D}_C[\boldsymbol{a}_1]-\mathcal{D}^j_S[\boldsymbol{a}_1]\right\}m_1\delta_{1} +(1\leftrightarrow 2)+\mathcal{O}(v^2),
\\
T^{ij}= & - v_1^{(i}\mathcal{D}^{j)}_S[\boldsymbol{a}_1]m_1\delta_{1} + (1\leftrightarrow 2)+\mathcal{O}(v^2).
\label{energymomentumofsource}
\end{aligned}
\end{align}
Again, the slow-motion approximation is achieved by discarding nonlinear velocity contributions.

In the following, angular brackets $\langle \dots\rangle$ denote a symmetric-trace-free (STF) projection of the respective indices. Furthermore, from the energy-momentum tensor \eqref{energymomentumofsource} it can be seen that $\ddot{T}^{ij}\sim\order(v^3,vm/r)$ and $\dot{T}^{ij}\sim\order(v^2,m/r)$ contribute at next-to-leading order. Considering this, the multipole moments of the post-Newtonian source reduce to \cite{Blanchet:2013haa}
\begin{align}\label{ILJL}
\begin{aligned}
\mathcal{I}_L= & \ \text{FP} \int d^3x\int_{-1}^1dz\bigg\{\delta_\ell(z)x_{\langle L\rangle}(T^{00}+T^{ij}\delta_{ij}) \\
 & -\frac{4(2\ell+1)}{(\ell+1)(2\ell+3)}\delta_{\ell+1}(z)x_{\langle a L\rangle}\dot{T}^{0a} \bigg\}, \\
\mathcal{J}_L= & \ \text{FP} \int d^3x\int_{-1}^1dz \ \tensor{\epsilon}{^a_b_{\langle i_\ell}}\bigg\{\delta_\ell(z)x_{L-1\rangle a}T^{0b}\bigg\}.
\end{aligned}
\end{align}
In these expressions, the energy-momentum tensor $T^{\mu\nu}=T^{\mu\nu}(\boldsymbol{x},\tilde{t}+zr)$ is a function of the extended time $\tilde{t}+zr$, which, together with the associated weighting function $\delta_\ell(z)$, takes the finite size of the source and the resulting time retardation into account. As argued above, at leading post-Newtonian order the finite-size-retardation vanishes since
\begin{align}
\int_{-1}^1dz\,  \delta_\ell(z)T^{\mu\nu}(\boldsymbol{x},\tilde{t}+zr)=T^{\mu\nu}(\boldsymbol{x},\tilde{t})+\order(v^2).
\end{align}

In principle, the total energy flux, as well as the gravitational wave modes, depend on all source multipole moments. However, as discussed below, only the mass- and current-quadrupole and -octopole contain leading PN information.
\begin{widetext}
\noindent In that context, the mass-type moments of the compact binary are 
\begin{subequations}
\begin{align}
\begin{split}
\mathcal{I}^{ij}= & \ m_1 z_1^{\langle ij\rangle}+\frac{4}{3}\left(2v_1^a S_1^b \tensor{\epsilon}{_a_b^{\langle i}}z_1^{j\rangle}-z_1^a S_1^b\tensor{\epsilon}{_a_b^{\langle i}}v_1^{j\rangle}\right)- \frac{1}{m_1}S_1^{\langle i}S_1^{j\rangle}+(1\leftrightarrow 2)+\order(v^2), \\
\mathcal{I}^{ijk}= & \ m_1 z_1^{\langle ijk\rangle} + \frac{3}{2}\left(3 v_1^a S_1^b \tensor{\epsilon}{_a_b^{\langle i}}z_1^{jk\rangle}-2 z_1^a S_1^b\tensor{\epsilon}{_a_b^{\langle i}}v_1^{j}z_1^{k\rangle} \right) - \frac{3}{m_1}S_1^{\langle i}S_1^jz_1^{k\rangle}- \frac{3}{2m_1^2} v_1^a S_1^b \tensor{\epsilon}{_a_b^{\langle i}}S_1^jS_1^{k\rangle} + (1\leftrightarrow 2)+\order(v^2), \\
%
%
%
\end{split}
\end{align}
and similarly, the current-type moments are
\begin{align}
\begin{split}
\mathcal{J}^{ij}= & \ m_1 z_1^a v_1^b\tensor{\epsilon}{_a_b^{\langle i}}z_1^{j\rangle}+\frac{3}{2}S_1^{\langle i}z_1^{j\rangle} +\frac{1}{m_1} v_1^a S_1^b\tensor{\epsilon}{_a_b^{\langle i}}S_1^{j\rangle}+(1\leftrightarrow 2)+\order(v^2), \\
\mathcal{J}^{ijk}= & \ m_1 z_1^a v_1^b\tensor{\epsilon}{_a_b^{\langle i}}z_1^{jk\rangle}+2S_1^{\langle i}z_1^{jk\rangle}+\frac{1}{m_1}\left(2 v_1^a S_1^b\tensor{\epsilon}{_a_b^{\langle i}}S_1^j z_1^{k\rangle}-z_1^a v_1^b\tensor{\epsilon}{_a_b^{\langle i}}S_1^j S_1^{k\rangle} \right)-\frac{2}{3m_1^2}S_1^{\langle i}S_1^j S_1^{k\rangle} + (1\leftrightarrow 2)+\order(v^2).
%
%
%
\end{split}
\end{align}
\label{multipolemoment}
\end{subequations}
\end{widetext}

\subsection{The total gravitational wave energy flux} \label{fluxsection}

The total GW energy flux can be directly obtained from the source multipole moments computed in the last section with the well-known relation \cite{Thorne:1980ru}
\begin{align}
\begin{aligned}
\mathcal{F}=\sum_{\ell=2}^{\infty}\bigg\{ & \frac{(\ell+1)(\ell+2)}{(\ell-1)\ell \ell! (2\ell+1)!!}\dot{\mathcal{U}}_L\dot{\mathcal{U}}^L\\
& +\frac{4\ell(\ell+2)}{(\ell-1)(\ell+1)!(2\ell+1)!!}\dot{\mathcal{V}}_L\dot{\mathcal{V}}^L\bigg\} .
\label{generalflux}
\end{aligned}
\end{align}
where the system's radiative moments $\mathcal{U}_L$ and $\mathcal{V}_L$ depend on the $\ell$-th time derivative of the source moments $\mathcal{I}_L$ and $\mathcal{J}_L$, as well as auxiliary source moments $W_L, X_L, Y_L$, and $Z_L$ as described in \cite{Blanchet:2013haa}. However, at leading post-Newtonian order we have simply $\mathcal{U}_L=\mathcal{I}_L^{(\ell)}$ and $\mathcal{V}_L=\mathcal{J}_L^{(\ell)}$, with $f^{(\ell)}(\tilde{t}):=d^\ell/dt^\ell f(t-R)$. In the spin-aligned configuration, the only time-dependent quantities in \eqref{multipolemoment} are $\bs{v}(\tilde{t})=v \bs{\lambda}(\tilde{t})$ and $\bs{r}(\tilde{t})=r \bs{n}(\tilde{t})$, where $\bs{\lambda}^{(\ell)}\sim\order(x^{3\ell/2})$ and $\bs{n}^{(\ell)}\sim \order(x^{3\ell/2})$. Taking also the spin contributions from $r=r_{\text{LO-S}^\infty}(x)$ into account, using $v=\omega r=x^{3/2}M^{-1} r(x)$, only the source quadrupole moments contribute at leading PN order. Therefore, expression \eqref{generalflux} reduces to 
\begin{align}
\mathcal{F}=\frac{1}{5}\dddot{\mathcal{I}}_{ij}\dddot{\mathcal{I}}^{ij}+\frac{16}{45}\dddot{\mathcal{J}}_{ij}\dddot{\mathcal{J}}^{ij}.
\end{align}
Carefully combining the individual leading PN order contributions at each spin order, the total gravitational wave energy flux of the binary black hole simplifies to
\begin{widetext}
\begin{align}
\begin{aligned}
\mathcal{F}_{\text{LO-S}^{\infty}}= \frac{\mu^2x^5}{M^2}\bigg[\frac{32}{5}- & \frac{8x^{3/2}}{5M}\Big\{8a_0+3\delta a_-\Big\}+\frac{2x^2}{5M^2}\Big\{32a_0^2+a_-^2\Big\}-\frac{4x^{7/2}}{15M^3}\Big\{16a_0^3+2a_0a_-^2+52\delta a_0^2a_-+\delta a_-^3\Big\} \\
& +\frac{2x^4a_0^2}{5M^4}\Big\{16a_0^2+a_-^2\Big\}+\frac{2a_0^2x^{11/2}}{15M^5}\Big\{64a_0^3+a_0a_-^2-68\delta a_0^2a_--3\delta a_-^3\Big\} \bigg].
\label{totalfluxresult}
\end{aligned}
\end{align}
\end{widetext}
Similar to the conserved energy, the total energy flux assumes the pattern described in \tablename{ \ref{PNorder}}, where in this case $A_{p>5}=0$. $\mathcal{F}_{\text{LO-S}^\infty}$ reproduces the results presented in \cite{Marsat:2014xea} up to cubic-in-spin effects. Again, the infinite sets of spin-induced multipolar interactions of the two black holes remarkably cancel out at higher than quintic-in-spin contributions. Hence, the total energy flux conveys the complete information about the spin effects at leading PN order in the first five terms of the spin expansion.

\subsection{Far zone gravitational wave modes $\bs{h}^{\bs{\ell m}}$} \label{modessection}

The angular distribution of the energy flux, as well as the frequencies, are encoded in the gravitational wave modes $h^{\ell m}$. In the proceedings, we follow the conventions of \cite{Kidder:2007rt}. As outlined above, we choose to describe the radiative dynamics in Cartesian coordinates on the background, with the center of mass of the BBH at the origin. Then, the defined spatial triad $\{\bs{n},\bs{\lambda},\bs{\ell}\}$ can be written as
\begin{align}
\begin{aligned}
\bs{n}= & \ (\cos \omega \tilde{t},\sin \omega \tilde{t},0),\\
\bs{\lambda}= & \ (-\sin \omega \tilde{t},\cos \omega \tilde{t},0),\\
\bs{\ell}= & \ (0,0,1).
\end{aligned}
\end{align}
Additionally, we define the radially outwards pointing 3-vector $\bs{N}$ by
\begin{align}
\bs{N}=(\sin \Theta \cos \Phi, \sin \Theta \sin \Phi, \cos \Theta),
\label{Nvector}
\end{align}
such that $\bs{R}:=R\bs{N}$ (see also \figurename{ \ref{plot}}, with $\omega \tilde{t}=\phi$). Furthermore, the STF spherical harmonics $\mathcal{Y}^{\ell m}_L$ are defined by $Y^{\ell m}(\Theta,\Phi)=\mathcal{Y}^{\ell m}_L N^L$, with the usual spherical harmonics $Y^{\ell m}(\Theta,\Phi)$.

The gravitational wave modes $h^{\ell m}$ are the projections of the polarization waveforms $h_+-i h_{\times}$ onto the spin weighted spherical harmonics $_{-2}Y^{lm}(\Theta,\Phi)$ with weight $s=-2$:
\begin{align}
h_+-i h_{\times}=\sum_{\ell =0}^\infty \sum_{m=-\ell}^\ell h^{\ell m} \ _{-2}Y^{\ell m}(\Theta,\Phi).
\label{polwaveform}
\end{align}
Making use of the Wigner $d$-function, the spin weighted spherical harmonics are 
\begin{align}
_{-s}Y^{\ell m}(\Theta,\Phi)=(-1)^s\sqrt{\frac{2\ell+1}{4\pi}}d^\ell_{m,s}(\Theta)e^{im\Phi},
\end{align}
with ${}_{0}Y^{\ell m}=Y^{\ell m}$, where 
\begin{align}
\nonumber& d^\ell_{m,s}(\Theta)= N_{\ell,m} \\
& \times \sum_{k=k_{\text{max}}}^{k_{\text{min}}}\frac{(-1)^k\left(\sin \Theta /2\right)^{2k+s-m}\left(\cos \Theta /2\right)^{2\ell+m-s-2k}}{k!(\ell+m-k)!(\ell-s-k)!(s-m+k)!},
\end{align}
with
\begin{align}
\begin{aligned}
N_{\ell,m}= & \ \sqrt{(\ell+m)!(\ell-m)!(\ell+s)!(\ell-s)!}, \\
k_{\text{max}}= & \ \max (0,m-s),\\
k_{\text{min}}= & \ \min (\ell+m,\ell-s).
\end{aligned}
\end{align}
Here $s$ is referred to as the spin weight, and $\ell$ and $m$ are the usual spherical-harmonic indices.  The far zone gravitational wave modes are generally given by
\begin{align}
h^{\ell m}=\frac{1}{\sqrt{2}R}\left[ \ \mathcal{U}^{\ell m}(\tilde{t})-i \mathcal{V}^{\ell m}(\tilde{t})\right],
\end{align}
with 
\begin{align}
\mathcal{U}^{\ell m}= & \ \frac{16 \pi}{(2\ell +1)!!}\sqrt{\frac{(\ell+1)(\ell+2)}{2\ell (\ell -1)}} \ \mathcal{U}^L\mathcal{Y}^{\ell m *}_L,\\
\mathcal{V}^{\ell m}= & - \frac{32\pi\ell}{(2\ell+1)!!}\sqrt{\frac{(\ell+2)}{2\ell(\ell+1)(\ell-1)}} \ \mathcal{V}^L\mathcal{Y}^{\ell m *}_L,
\end{align}
where $^*$ denotes complex conjugation. 

Using the system multipole moments given in \eqref{multipolemoment}, with their time derivatives as discussed in Sec.~\ref{fluxsection}, allows us to compute the modes $h^{\ell m}$ for $\ell=2$ and $\ell=3$, with the leading-PN-order terms at each order in spin for all those modes.  (Note that our level of approximation would in fact allow us to do this for arbitrary $\ell$, given extensions of \eqref{multipolemoment} to the hexadecapole moments and beyond, which could be computed from (\ref{ILJL}) just as we did for the quadrupoles and octupoles.)  Defining the symmetric mass ratio,
\begin{equation}
\nu=\frac{\mu}{M}=\frac{m_1m_2}{M^2},
\end{equation}
and factorizing the modes according to
\begin{align}
h^{\ell m}_{\text{LO-S}^{\infty}}=\sqrt{\pi}\nu\frac{M}{ R} \hat{h}^{\ell m} e^{-i m\omega \tilde{t}},
\end{align}
the structures of the results (for general $\ell$) are as follows, 
\begin{widetext}
\begin{subequations}
\begin{align}
\ell-m\;\;\textrm{even}\;:\qquad \hat h^{\ell m}&=x^{\ell/2}\left(\sum_{p\textrm{(even)}=0}^\infty x^p+\sum_{p\textrm{(odd)}=1}^\infty x^{p+1/2}\right)\sum_{q=0}^p C^{\ell m}_{pq}\frac{a_0^{p-q}a_-^q}{M^p_{\phantom{0}}}\Big(1+\order(x)\Big),
\\
\ell-m\;\;\textrm{odd}\;:\qquad \hat h^{\ell m}&=x^{\ell/2}\left(\sum_{p\textrm{(even)}=0}^\infty x^{p+1/2}+\sum_{p\textrm{(odd)}=1}^\infty x^{p}\right)\sum_{q=0}^p C^{\ell m}_{pq}\frac{a_0^{p-q}a_-^q}{M^p_{\phantom{0}}}\Big(1+\order(x)\Big),
\end{align}
\end{subequations}
where the coefficients $C^{\ell m}_{pq}$ depend only on the dimensionless mass ratios $\nu$ and $\delta$.  Note that the pattern of relative orders for the case when $\ell-m$ is even is as in \tablename{ \ref{PNorder}}, while the pattern for the case when $\ell-m$ is odd differs.

Dropping here the next-to-leading-order correction factors $(1+\order(x))$, for $\ell=2,3$, we find $\hat{h}^{20}=0=\hat{h}^{30}$ and
\begin{subequations}
\begin{align}
\hat{h}^{22}= & \ -\frac{8 x}{3\sqrt{5}}\Bigg\{ 3\left(1+x^2\frac{a_0^2}{M^2}\right)
\nonumber\\
& \ \qquad\qquad
-x^{3/2}\frac{3a_0+\delta a_-}{M}+2x^{7/2}\frac{a_0^2(a_0- \delta a_-)}{M^3}\Bigg\},
\\
\hat{h}^{21}= & \ -\frac{2i  x^{3/2}}{3\sqrt{5}}\Bigg\{\left(1+x^2\frac{a_0^2}{M^2}\right)^{-1/2}\left[4\delta +2x^2\frac{4\delta a_0^2+2a_0 a_-+\delta a_-^2}{M^2} +x^4\frac{a_0^2(4 \delta a_0^2-a_0a_-+3\delta a_-^2 )}{M^4}\right]
\nonumber \\ 
& \qquad\qquad\quad -\left(1+x^2\frac{a_0^2}{M^2}\right)^{1/2} 6x^{1/2}\frac{a_-}{M}\Bigg\},
\\
\hat{h}^{33}= & \ \frac{3i}{4}\sqrt{\frac{6}{7}}  x^{3/2}
\Bigg\{
4\delta\left(1+x^2\frac{a_0^2}{M^2}\right)^{3/2} 
\nonumber \\
&\qquad\qquad\qquad+\left(1+x^2\frac{a_0^2}{M^2}\right)^{1/2}\left[ -x^{3/2}\frac{7\delta a_0+(1-10\nu)a_-}{M}
 +3 x^{7/2}\frac{a_0^2(\delta a_0-(1-6\nu)a_-)}{M^3}\right]
\Bigg\}, 
\\
\hat{h}^{32}= & \ -\frac{4  x^{2}}{9\sqrt{7}}\Bigg\{6(1-3\nu)+x^2\frac{(5-36\nu)a_0^2+6\delta a_0a_-+(1-16\nu)a_-^2}{M^2}+3x^4\frac{a_0^2((3-6\nu)a_0^2-2\delta a_0a_-+(1-8\nu)a_-^2)}{M^4}
\nonumber \\ 
& \qquad\qquad\;\;
+\left(1+x^2\frac{a_0^2}{M^2}\right)6x^{1/2}\frac{a_0-\delta a_-}{M}
\Bigg\},
\\
\hat{h}^{31}= & \ -\frac{i}{36}\sqrt{\frac{2}{35}} x^{3/2}\Bigg\{ 12\left(1+x^2\frac{a_0^2}{M^2}\right)^{1/2}\left[ \delta+x^2\frac{a_0(\delta a_0-4 a_-)}{M^2}\right]
\nonumber\\
&\qquad\qquad\qquad\quad+\left(1+x^2\frac{a_0^2}{M^2}\right)^{-1/2}\bigg[
3x^{3/2}\frac{\delta a_0-(9-26\nu)a_-}{M}+2x^{7/2}\frac{18\delta a_0^3-(13-90\nu)a_0^2a_-+8\delta a_0a_-^2+3a_-^3}{M^3}
\nonumber\\
&
\qquad\qquad\qquad\qquad\qquad+3x^{11/2}\frac{a_0^2(11\delta a_0^3-(13-34\nu)a_0^2a_-+8\delta a_0a_-^2+2a_-^3)}{M^5}\bigg]
\Bigg\},
\end{align}
\label{hlmmodes}
\end{subequations}
\end{widetext}
with the negative-$m$ modes given by $h^{\ell,-m}=(-1)^\ell h^{\ell m*}$. Note that the spin expansion of the even-$m$ modes terminates at a finite order, while the odd-$m$ modes have contributions at all orders in spin (once all factors are expanded in spin).  Considering the polarization waveform \eqref{polwaveform}, with the full sum over modes, and taking the leading PN orders (lowest powers of $x$) at each order in spin, as opposed to doing this for each mode separately, we see that only (parts of) the $\ell=2,3$ results given here contribute.

\section{Conclusion}

We determined the binding energy, the gravitational wave modes, and total energy flux emitted by a spinning nonprecessing binary black hole in quasicircular motion at leading post-Newtonian orders at all orders in spin.  Our results include contributions of arbitrarily large PN order, counting in $1/c^2$.  In particular, we obtained for the first time the quartic-in-spin contributions to the 4PN waveform and total energy flux, along with all higher-order-in-spin contributions at the corresponding leading PN orders. Remarkably, the binding energy, the total energy flux, as well as the even-in-$m$ gravitational wave modes only contain a finite number of nonzero contributions in their spin expansions at leading post-Newtonian order, when expressed in terms of the circular-orbit frequency. For instance, we showed that previously found cubic-in-spin results \cite{Marsat:2014xea} for the energy-frequency relationship are in fact valid to all orders in spin without additional corrections.

Conversely, the modes where all powers in spin appear are nevertheless rather compact, which can be used to improve the resummation of modes, e.g., in the synergetic EOB waveform model \cite{Pan:2010hz}. Though our results are only valid for aligned spins, they can still be used to approximate waveforms from precessing binaries \cite{Pan:2013rra}. We leave the investigation of precessing systems or analysis of possibly similar resummations of all spin orders at next-to-leading post-Newtonian order to future work. 

\begin{acknowledgments}
We are grateful to Sylvain Marsat for useful discussions and for sharing his unpublished results for the GW modes at cubic order in spin, and to Alessandra Buonanno for her helpful comments on an earlier version of this paper. We also thank the referee for many useful comments.  This work was presented in the bachelor's thesis of N.S. at the Humboldt-Universität zu Berlin. N.S. thanks the Albert Einstein Institute for its hospitality during the completion of this project.
\end{acknowledgments}

\appendix
\section{Noether charges from Poincar\'e symmetry}\label{app:Noe}

Here we discuss the application of Noether's theorem to determine conserved quantities for our BBH system at leading PN order, following, e.g., the treatment in 
\cite{deAndrade:2000gf}.

Our BBH effective action with Lagrangian \eqref{Leff}, valid for generic orbits and spin orientations, before specializing to the center-of-mass frame, is of the form
\begin{equation}
S=\int dt\,\mathcal L(z_A^i,v_A^i,\dot v_A^i,\Lambda_A^{Ki},\dot \Lambda_A^{Ki},S_A^i)=\int dt\,\mc L(\psi),
\end{equation}
where $A=1,2$ labels the two BHs, recalling that $K=1,2,3$ is a spatial body-fixed frame index, and that $v_A^i=\dot z_A^i$, and using $\psi$ to collectively denote all of the degrees of freedom and their time derivatives.  The generalized Euler-Lagrange equations resulting from varying the action with respect to $z_A^i(t)$, $\Lambda_A^{Ki}$, and $S_A^i$ are
\begin{align}
0= \frac{\partial \mathcal L}{\partial {z}_A^i}-\frac{d}{dt} \frac{\partial \mathcal L}{\partial {v}^i_A}+\frac{d^2}{dt^2}\frac{\partial \mathcal L}{\partial \dot{{v}}^i_A}.
\label{generalELeq}
\end{align}
and
\begin{align}
0=\frac{\partial\mathcal{L}}{\partial \Lambda_A^{Ki}}-\frac{d}{dt} \frac{\partial \mathcal L}{\partial \dot{\Lambda}_A^{Ki}},
\qquad
0= \frac{\partial \mathcal L}{\partial {S}^i_A}.
\label{spinELeq}
\end{align}
Under a general infinitesimal transformation of all of the degrees of freedom, $\psi\to\psi+\delta\psi$, the change in the Lagrangian is given by
\begin{align}
\begin{aligned}
\delta\mathcal{L} &=
\frac{\doe\mc L}{\doe z_A^i}\delta z_A^i
+\frac{\doe\mc L}{\doe v_A^i}\delta v_A^i
+\frac{\doe\mc L}{\doe \dot v_A^i}\delta \dot v_A^i
\\
&\quad+\frac{\doe\mc L}{\doe \Lambda_A^{Ki}}\delta \Lambda_A^{Ki}
+\frac{\doe\mc L}{\doe \dot\Lambda_A^{Ki}}\delta \dot\Lambda_A^{Ki}
+\frac{\doe\mc L}{\doe S_A^i}\delta S_A^i
\\
&=\frac{d}{dt}\left(p^A_i\delta z_A^i+q^A_i\delta v_A^i
+\frac{\doe\mc L}{\doe\dot\Lambda_A^{Ki}}\delta\Lambda_A^{Ki}\right),
\end{aligned}
\end{align}
where sums over $A$ are implied, where the last line has used the equations of motion \eqref{generalELeq} and \eqref{spinELeq}, and where we define the generalized momenta
\begin{align}\label{pq}
p^A_i:=\frac{\doe\mc L}{\doe v_A^i}-\dot q^A_i,\qquad q^A_i:=\frac{\doe\mc L}{\doe\dot v_A^i}.
\end{align}
If the transformation is a symmetry of the action, leaving the action invariant, then the change in the Lagrangian must be a total time derivative of some function $f$ (without using the equations of motion),
\begin{equation}\label{dfdt}
\delta \mc L=\frac{d}{dt}f(\psi,\delta\psi),
\end{equation}
and it follows that the Noether charge
\begin{align}\label{bigQgen}
Q=\sum_A\left(p^A_i\delta z_A^i+q^A_i\delta v_A^i
+\frac{\doe\mc L}{\doe\dot\Lambda_A^{Ki}}\delta\Lambda_A^{Ki}\right)-f
\end{align}
is a conserved quantity, $dQ/dt=0$.  

We expect our action to be invariant under (the leading-PN-order limit of) the full Poincar\'e group of spacetime symmetries, leading to the full set of ten conserved Poincar\'e charges.  We will confirm here that this is so and derive expressions for the charges.

Let us first specialize to the case, as in (\ref{Leff}), in which the Lagrangian depends on the body-fixed triads $\Lambda_A^{Ki}$ only through the terms $\Omega^A_k S_A^k=\frac{1}{2}\epsilon_i{}^j{}_k\Lambda_A^{Ki}\dot\Lambda^A_{Kj}S_A^k$ in $\mc L$.  The Noether charge (\ref{bigQgen}) can then be written as
\begin{align}\label{bigQ}
Q=\sum_A\left(p^A_i\delta z_A^i+q^A_i\delta v_A^i
+\frac{1}{2}\epsilon^i{}_{jk}S_A^k\Lambda^A_{iK}\delta\Lambda_A^{Kj}\right)-f.
\end{align}

Now, under a time translation, we have $\psi(t)\to\psi(t+\epsilon)$ with $\epsilon$ constant so that $\delta\psi=\epsilon\dot\psi$, and it is clear that $f=\epsilon \mc L$ in (\ref{dfdt}).  We then find from (\ref{bigQ}) that $Q=\epsilon E$ where the conserved total energy is
\begin{align}\label{NoeE}
E=\sum_A\left(p^A_iv_A^i+q^A_i\dot v_A^i
+\Omega_i^AS_A^i\right)-\mc L.
\end{align}

Under a spatial translation, we have $\delta z_A^i=\epsilon^i$ with $\epsilon^i$ being a constant vector, and all the other $\delta\psi$ are zero.  Because the Lagrangian (\ref{Leff}) depends on $z_A^i$ only through $r^i=z_1^i-z_2^i$, the Lagrangian itself is invariant, and $f=0$.  We then find $Q=\epsilon^i P_i$ where the conserved total momentum is
\begin{equation}
P_i=\sum_Ap_i^A.
\end{equation}

Under a spatial rotation, we have $\delta \psi^i=\epsilon^i{}_{jk}\theta^j \psi^k$ where $\theta^i$ is a constant vector, for $\psi^i$ being each of the vectors $z_A^i$, $\Lambda_A^{Ki}$ (for each $K$), $S_A^i$, and their time derivatives.  The Lagrangian is also invariant under this transformation, and $f=0$.  Using $\Lambda_A^{Ki}\Lambda^A_{Kj}=\delta^i_j$, we find $Q=\theta^iJ_i$ where the conserved total angular momentum is
\begin{equation}\label{NoeJ}
J_i=\sum_A\left[\epsilon_{ij}{}^k\left(z_A^jp^A_k+v_A^jq^A_k\right)+S^A_i\right].
\end{equation}

Finally, we consider a boost, which at leading PN order reduces to a Galilean transformation, with $\delta z_A^i=w^i t$ and $\delta v_A^i=w^i$, where $w^i$ is a constant vector, and with all the other $\delta\psi$ being zero.  The change in the Lagrangian (\ref{Leff}) is given by
\begin{align}
\begin{aligned}
\delta \mc{L}&=w_i\sum_A\left(m_Av_A^i+\frac{1}{2}\epsilon^i{}_{jk}\dot v_A^jS_A^k\right)
\\
&=\frac{d}{dt}\left[w_i\sum_A\left(m_Az_A^i+\frac{1}{2}\epsilon^i{}_{jk} v_A^jS_A^k\right)\right]+\mathrm{NLO},
\label{dLboost}
\end{aligned}
\end{align}
where NLO indicates (minus) the next-to-leading-PN-order terms that would arise from the time derivative acting on $S_A^i$; these terms are all suppressed by a factor of $\epsilon_\mathrm{PN}\sim v^2\sim M/r$ relative to the last term in the first line, as can be verified from the expression for $\dot S_A^i$ obtained from \eqref{spinELeq} with \eqref{Leff}.\footnote{We admit that this argument for dropping the $\dot S_A^i$ contributions in \eqref{dLboost} goes against the fact that one should not use the equations of motion in showing that $\delta\mc L$ is a total time derivative, but this is ultimately inconsequential for the result, as the  dropped corrections to the conservation law are still at next-to-leading order.  One can verify that the $K^i$ in \eqref{KGP} indeed satisfies $\dot K^i=0$ as a consequence of the equations of motion, working consistently at leading PN order, for generic orbits and spin orientations.}   Thus, for the boost, the $f$ in (\ref{dfdt}) and (\ref{bigQ}) is given by the quantity in square brackets in \eqref{dLboost}.  From (\ref{bigQ}), we find $Q=-w_iK^i$ where the conserved quantity arising from the boost symmetry is
\begin{align}
\begin{aligned}\label{KGP}
K^i&=G^i-P^it,
\\
G^i&=\sum_A\left(m_Az_A^i+\epsilon^i{}_{jk}v_A^jS_A^k\right),
\end{aligned}
\end{align}
where $G^i$ is the system's mass-dipole vector.  Here, we have used the explicit expression for $q_A^i$ resulting from (\ref{pq}) and (\ref{Leff}).

Because $P^i$ is conserved, we can consistently choose a coordinate frame in which $P^i=0$, a center-of-momentum frame.  In such a frame, the conservation of $K^i$ implies that $\dot G^i=0$, and we can then further specialize the coordinates so that $G^i=0$, defining the center-of-mass frame.  
Equations \eqref{COMframe} employed in the main text result from setting $G^i=0$ in \eqref{KGP} and solving for $z_1^i$ and $z_2^i$ in terms of $r^i=z_1^i-z_2^i$.

The expressions \eqref{conservedenergy} for $E$ and \eqref{conservedanglmoment} for $J$ in the main text result from evaluating (\ref{NoeE}) and (\ref{NoeJ}), using (\ref{pq}) and (\ref{Leff}), while specializing to the aligned-spin, circular-orbit configuration, and using the center-of-mass-frame relations \eqref{COMframe}, working consistently at leading PN order.

\section{Action of $\bs{\mathcal{D}}_{\bs{C}}\bs{[}\boldsymbol{a}\bs{]}$ and $\bs{\mathcal{D}}^{\bs{i}}_{\bs{S}}\bs{[}\boldsymbol{a}\bs{]}$ on $\bs{r}^{\bs{-1}}$} \label{appendixB}

In the following, the action of the differential operators $\mathcal{D}_C[\boldsymbol{a}]$ and $\mathcal{D}^i_S[\boldsymbol{a}]$, defined in \eqref{diffopC} and \eqref{diffopS}, respectively, on $r^{-1}=|\boldsymbol{z}_1-\boldsymbol{z}_2|^{-1}$ are presented in detail. Recall from Sec.~\ref{eomsection} that the vectors $\{ \bs{n}, \boldsymbol{\lambda}, \boldsymbol{\ell}\}$ pose a positively oriented triad (i.e., $\boldsymbol{n}\times\boldsymbol{\lambda}=\boldsymbol{\ell}$) on the spatial slices of spacetime, where $\bs{r}=r\bs{n}$. Let us assemble the tools first \cite{poisson14}. In this appendix, we use $\partial_i=\partial/\partial z_1^i$; therefore, the simple identity
\begin{align}
\partial_M r^{-1}=(-1)^m (2m-1)!!\frac{n_{\langle M\rangle}}{r^{m+1}},
\label{partialofz}
\end{align}
where $\langle \dots \rangle$ indicates a symmetric trace-free (STF) projection, holds. Let $P_\ell(x)$ be the $\ell$th Legendre polynomial, then the STF contraction of two arbitrary unit vectors $\boldsymbol{h}$ and $\boldsymbol{h}'$, with relative angle $\alpha=\boldsymbol{h}\cdot\boldsymbol{h}'$, can be shown to be
\begin{align}
h'_{\langle M\rangle}h^{\langle M\rangle}=\frac{m! P_m(\alpha)}{(2m-1)!!}.
\end{align}
Lastly, we recall that 
\begin{align}
P_m(0) =
\begin{cases}
\frac{(-1)^n}{4^n}{{2n}\choose{n}},  & m=2n, \\
0, & \text{otherwise,}
\end{cases}
\label{legendrepolynomialzero}
\end{align}
with binomial coefficient ${{k}\choose{l}}$. Because of spin alignment, we only need to consider arbitrary vectors $\boldsymbol{a}:=a\boldsymbol{\ell}$, where $a:=\boldsymbol{a}\cdot \boldsymbol{\ell}$. Thus, using \eqref{diffopC}, we find
\begin{align}
\mathcal{D}_C[\boldsymbol{a}]r^{-1}= \sum_{m =0}^{\infty} (-1)^m a^{2m}\frac{P_{2m}(\bs{n}\cdot\boldsymbol{\ell})}{r^{2m+1}}.
\end{align}
Since $P_{2m}(\bs{n}\cdot\boldsymbol{\ell})$ is independent of $\bs{r}$, the circular motion restriction can be used at this stage (even when additional derivatives need to be taken). We obtain, after making use of \eqref{legendrepolynomialzero} and resumming the resulting series,
\begin{align}
\begin{aligned}
\mathcal{D}_C[\boldsymbol{a}]r^{-1} & =\sum_{m =0}^{\infty}{{2m} \choose{m}}\left(\frac{a}{2}\right)^{2m} r^{-2m-1}\\
& = \frac{1}{\sqrt{r^2-a^2}}.
\label{coshonr}
\end{aligned}
\end{align}
Any additional spatial differentiation of this result with $\partial$ can be done simply by using the first line of \eqref{coshonr}. For instance, the expression $\partial^j \mathcal{D}_C[\boldsymbol{a}]r^{-1}$ can be computed by noting $\partial_i r=n_i$. Using \eqref{partialofz}, and resumming, we acquire
\begin{align}
\partial^j \mathcal{D}_C[\boldsymbol{a}]r^{-1}= \frac{-r^j}{(r^2-a^2)^{3/2}}.
\end{align}

The action of $\mathcal{D}^i_S[\boldsymbol{a}]$ on $r^{-1}$ is constructed similarly. Using \eqref{diffopS}, the differential operator is given by
\begin{align}
\begin{aligned}
\mathcal{D}^i_S[\boldsymbol{a}]r^{-1}= & \ a^j\tensor{\epsilon}{_j^i^d}\sum_{m =0}^{\infty}(-1)^m a^{2m}\frac{n_d P_{2m}(\bs{n}\cdot\boldsymbol{\ell})}{r^{2m+2}} \\
= & \ -\frac{a \lambda^i}{r\sqrt{r^2-a^2}} .
\end{aligned}
\end{align}
For example, applied to the case $v^j\partial_j \mathcal{D}^i_S[\boldsymbol{a}]r^{-1}$, using \eqref{partialofz} and resumming, we find
\begin{align}
v^j\partial_j \mathcal{D}^i_S[\boldsymbol{a}]r^{-1}=\frac{var^i}{r^3\sqrt{r^2-a^2}},
\end{align}
where $v=|\bs{v}|$. The remaining necessary expressions for \eqref{zeom} can be gathered in this way. We find:
\begin{align}
\partial^jv_i\mathcal{D}^i_S[\boldsymbol{a}]r^{-1}= & - r^j\frac{v a}{r^3}\frac{a^2-2r^2}{(r^2-a^2)^{3/2}}, \\
v^i\sigma^{*k}\tensor{\epsilon}{_k_l^j}\partial_i\partial^l\mathcal{D}_C[\boldsymbol{a}]r^{-1}= & \ r^j\frac{v\sigma^*}{r}\frac{1}{(r^2-a^2)^{3/2}}.
\end{align}
Where, as defined in the text, $\sigma^*=\boldsymbol{\ell}\cdot\boldsymbol{\sigma}^*$.

%

\end{document}